\documentclass[aps,prl,11pt,reprint,superscriptaddress]{revtex4-1}

\usepackage{textcomp}
\usepackage[normalem]{ulem}
\usepackage{amsmath}
\usepackage{textcomp}
\usepackage{amssymb}
\usepackage{scalerel}
\usepackage{graphicx,color}
\usepackage{braket}
\usepackage{footnote}
\usepackage{placeins}
\usepackage{float}
\usepackage{booktabs}
\usepackage{xcolor}
\usepackage{titlesec}
\titlespacing\section{0pt}{12pt plus 4pt minus 2pt}{0pt plus 2pt minus 2pt}
\titlespacing\subsection{0pt}{12pt plus 4pt minus 2pt}{0pt plus 2pt minus 2pt}
\titlespacing\subsubsection{0pt}{24pt plus 4pt minus 2pt}{0pt plus 2pt minus 2pt}
\usepackage[top=0.8in, bottom=0.8in, left=1.2in, right=1.2in]{geometry}
\usepackage{multirow}
\usepackage{amsbsy}
\usepackage{array}
\usepackage{soul}

\mathchardef\mhyphen="2D

\usepackage{braket}

\begin{document}
\title{Hong-Ou-Mandel interference of polarization qubits stored in independent room-temperature quantum memories}

\author{Sonali Gera}
\altaffiliation{S.G. and C.W. contributed equally to this work.}
\affiliation{Department of Physics and Astronomy, Stony Brook University, Stony Brook, NY 11794, USA}

\author{Chase Wallace}
\altaffiliation{S.G. and C.W. contributed equally to this work.}
\affiliation{Department of Physics and Astronomy, Stony Brook University, Stony Brook, NY 11794, USA}

\author{Mael Flament}
\affiliation{Qunnect Inc., 141 Flushing Av. Suite 1110, Brooklyn, NY, 11205, USA}

\author{Alessia Scriminich}
\affiliation{Department of Information Engineering, University of Padova, Via Gradenigo 6b, 35131 Padova, Italy}

\author{Mehdi Namazi}
\affiliation{Qunnect Inc., 141 Flushing Av. Suite 1110, Brooklyn, NY, 11205, USA}

\author{Youngshin Kim}
\affiliation{Department of Physics and Astronomy, Stony Brook University, Stony Brook, NY 11794, USA}

\author{Steven Sagona-Stophel}
\affiliation{Department of Physics and Astronomy, Stony Brook University, Stony Brook, NY 11794, USA}

\author{Giuseppe Vallone}
\affiliation{Department of Information Engineering, University of Padova, Via Gradenigo 6b, 35131 Padova, Italy}

\author{Paolo Villoresi}
\affiliation{Department of Information Engineering, University of Padova, Via Gradenigo 6b, 35131 Padova, Italy}

\author{Eden Figueroa}\email[corresponding author: \\]{eden.figueroa@stonybrook.edu}
\affiliation{Department of Physics and Astronomy, Stony Brook University, Stony Brook, NY 11794, USA}
\affiliation{Brookhaven National Laboratory, Upton, NY, 11973, USA}

\begin{abstract}
\textbf{Quantum repeater networks require independent quantum memories capable of storing and retrieving indistinguishable photons to perform high-repetition entanglement swapping operations. The ability to perform these coherent operations at room temperature is of prime importance to the realization of scalable quantum networks. We perform Hong-Ou-Mandel (HOM) interference between photonic polarization qubits stored and retrieved from two sets of independent room-temperature quantum memories. We show a steady improvement in memory parameters and visibilities, culminating in a high quantum memory HOM visibility of 43\%, compared to the 48\% no-memory limit of our set-up. These results lay the groundwork for future applications using large-scale memory-assisted quantum networks.}
\end{abstract}

\maketitle
\newpage
\section{\large{I.  Introduction}}\label{sec:intro}

Quantum technologies have the promise of achieving several advantages in the fields of computation, metrology, and security \cite{Awschalom2021}. The development of a large-scale quantum internet \cite{Wehner2018} is an encouraging route to attain quantum advantage using distributed quantum systems. To this end, elementary quantum networks have already been deployed, capable of distributing entanglement over tens to hundreds of kilometers \cite{Luo2020,vanLeent2022,Neumann2022}. However, developing more robust and accessible long-distance entanglement distribution networks brings the challenge of mitigating losses. To sidestep the inherent non-amplifiability of quantum signals, one can harness entanglement swapping \cite{JWP1998}, the basis of two-way quantum repeaters (QRs) \cite{DLCZ2001}, thereby extending the range over which quantum information can be distributed. Quantum memories (QMs) \cite{Lvovsky2009} play a crucial role in these repeaters \cite{Azuma2023}, serving as temporary storage units capable of heralding and coordinating qubit arrival at Bell-state measurement stations, significantly increasing the probability of successful entanglement swapping. While many types of QMs exist, room-temperature atomic ensembles provide a particularly appealing platform to develop scalable photonic quantum repeaters \cite{Peng2022,Dideriksen2021}, the backbone for a user-defined entanglement-based quantum internet.\\

Currently, there are two types of quantum repeater network implementations with a degree of maturity. Type I is based upon the DLCZ protocol \cite{DLCZ2001}, wherein entanglement is generated by the interference of photons generated in quantum registers. While this method heralds the creation of entanglement between two registers, it has an inherently low rate \cite{Moehring2007}. Whereas in type II repeaters, entanglement is generated independently of the registers, and photons from separate entangled pairs are stored in a minimum of four in-out QMs after propagation \cite{Lloyd2001}. Once there, their presence is heralded \cite{Bhaskar2020} and the intermediary photon's entanglement is swapped via a Bell-state measurement (BSM). By interfacing fast entanglement sources and fast quantum memories, rates can become several orders of magnitude higher than type I \cite{Muralidharan2016}. Recent studies have demonstrated this potential of integrating high-duty-cycle room-temperature atomic quantum memories with fast entanglement sources \cite{Peng2022, Jin2021, Walmsley2023, Buser2022, Guodong2023}.\\

In the context of a type II repeater \cite{Rakonjac2021}, for the swapping of entanglement to be successful, retrieved qubits from the in-out memories must maintain the same quantum state as their inputs, such as polarization \cite{Namazi2017, Namazi2017_2, Qunnect2022} or time bin. The retrieved qubits must remain nearly indistinguishable from each other across all other crucial degrees of freedom, such as temporal and spectral profiles, thus the memories' performance must also remain nearly identical across those degrees of freedom as well as have similar overall efficiencies. A Hong-Ou-Mandel (HOM) interference measurement \cite{Mandel1987} on stored and released qubits \cite{Jin2013} provides a method to verify this identical behavior of memories. Despite previous work focusing on type I \cite{Hansen2022, Dounan2021} quantum repeaters, an experimental demonstration of HOM interference between two photonic qubits retrieved from type II quantum memory systems—a significant stride toward demonstrating type II memory assisted entanglement swapping—remains to be shown.\\

Here, we demonstrate, for the first time to our knowledge, HOM interference between two qubits stored and retrieved from independent type II quantum memory systems. In addition to this, we demonstrate this foundational measurement across four QMs, pairwise, in the configuration of that of a type II QR. We show that, in the limit of high signal-to-background ratio (SBR), the storage and retrieval of photonic qubits in our \cite{Namazi2017} warm Rb vapor quantum memories does not significantly degrade HOM visibility. We also show the detrimental effects of the background generated during the retrieval of single-photon-level qubits and describe a model quantifying the effect of this background on the HOM visibility. We also present a road map toward demonstrating even higher visibilities in HOM interference experiments using room-temperature quantum memories, laying the groundwork for memory-assisted BSMs of qubits and the first instance of a type II quantum repeater.\\

We present three sets of experiments. First, in section II, two phase-independent sources create coherent pulses at few-photon-level that serve as polarization qubit inputs to our first set of two independent room-temperature quantum memories (Gen I \cite{Namazi2017}). As a first test, HOM interference is observed between retrieved qubits by changing the relative polarizations of qubits retrieved from the memory. Second, in section III, the input qubits are attenuated to contain on average approximately a single photon per pulse. HOM interference is observed by varying the qubit generation time while maintaining a constant storage time. This allows us to investigate the effect of background on memory HOM visibility. Third, in section IV, a second set of quantum memories (Gen II \cite{Qunnect2022}) is employed with an approximate five-fold increase in the SBR. This time, HOM interference is observed by keeping the qubit generation time constant and varying the storage time, resulting in a significant increase in visibility.

\section{\large{II.  Gen I QM HOM interference with few-photon-level polarization qubits}} \label{sec:QubitHOM}

\begin{figure*}[!ht]
\vspace{1.5mm}
\centering
\includegraphics[width=0.98\textwidth]{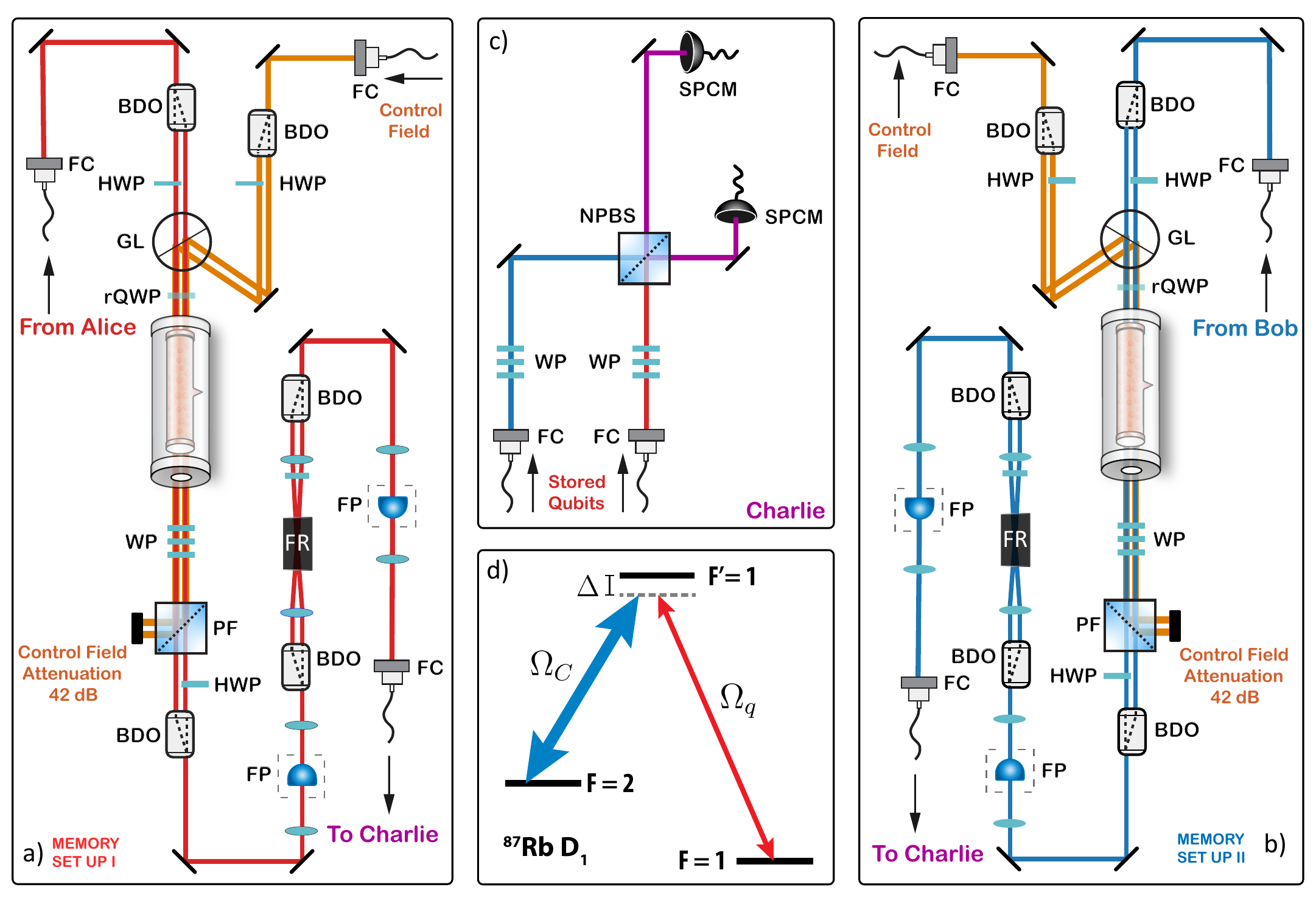}
\caption{\textbf{QM HOM Intereference set-up: four quantum light-matter interfaces forming two independent quantum memories for polarization qubits}. 
\textbf{(a,b)} Quantum memory setup. After fiber propagation, the pulses are stored in two independent dual-rail atomic vapor quantum memories, consisting of the light-matter interfaces, polarization, and frequency filtering stages; \textbf{(c)} The outputs of the quantum memories are made to interfere at a 50:50 non-polarizing beamsplitter; \textbf{(d)} Atomic level scheme for EIT storage in the $^{87}$Rb D$_1$ line. $\Omega_C$ represents the strong control field, and $\Omega_q$ represents our weak probe/qubit. They are red detuned from the $F'=1$ transition by $\Delta=400$MHz. \textbf{FC} Fiber Collimator, \textbf{BDO} Beam Displacment Optics, \textbf{HWP} Half Wave Plate, \textbf{WP} Waveplate Combination, \textbf{rQWP} removable Quarter Wave Plate, \textbf{GL} Glan Laser Polarizer, \textbf{PF} Polarization Filtering, \textbf{FP} Fabry Perot Etalon, \textbf{FR} Faraday Rotator, \textbf{NPBS} Non Polarizing Beam Splitter, \textbf{SPCM} Single Photon Counting Modules}
\label{fig:layout}
\label{fig:MemoryDiagram}
\end{figure*}

Two independent sources of polarization qubits, herein called Alice and Bob, serve as inputs to two independent room-temperature quantum memories. The memories rely on electromagnetically induced transparency (EIT) in a lambda scheme to store and retrieve qubit pulses in the D$_1$ line of warm $^{87}$Rb vapor, as shown in Fig. \ref{fig:MemoryDiagram}d. This is adapted in a dual rail configuration to store polarization qubits with high fidelity \cite{Namazi2017,Kupchak2015} (Fig. \ref{fig:MemoryDiagram}a,b). For our first set of experiments, we use two first-generation QMs to store and retrieve $400$ns wide qubits. After retrieval, the qubits are made to interfere on a 50:50 beamsplitter, shown in Fig. \ref{fig:MemoryDiagram}c. \\

\begin{figure}[!ht]
  \vspace{-2mm}
  \centering
  \includegraphics[width=1\columnwidth]{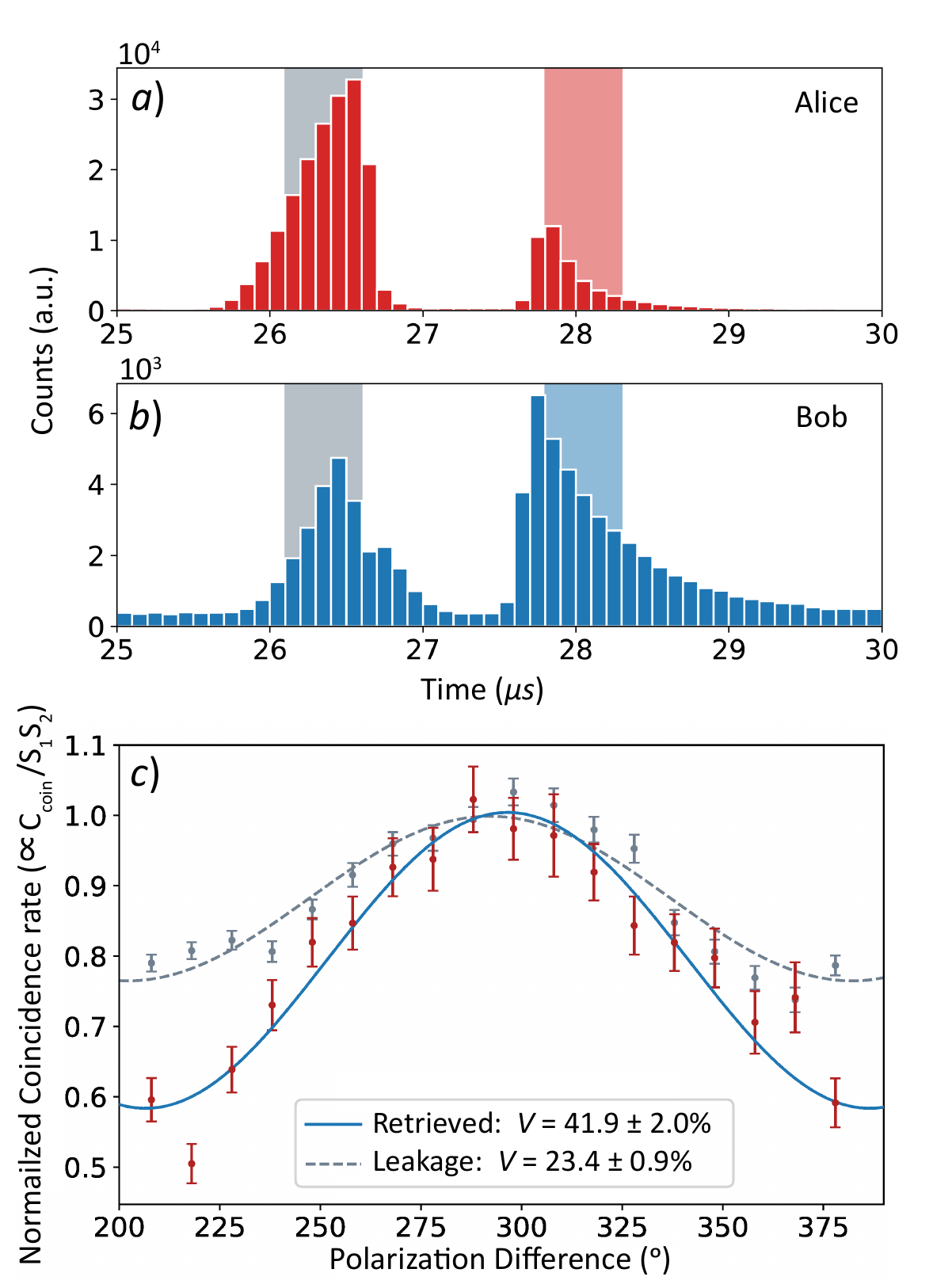}
  \caption{\footnotesize{\textbf{HOM interference of qubits retrieved from two Gen I quantum memories with few photon level inputs}}. \textbf{(a)} Histograms of single-photon detection events showing the temporal profile of the stored qubits in Alice's memory. The gray shaded region indicates the non-stored part of the qubits and the red shaded region indicates the part of the pulse that was stored and retrieved which contains about 0.024 photons per pulse. \textbf{(b)} Stored part of Bob contains about 0.017 photons per pulse. The gray and blue shaded regions indicate the leakage and stored regions of interest respectively. \textbf{(c)} The two detector coincidence rate is measured by varying the relative polarization of the retrieved pulses. We show the normalized coincidences after HOM interference for pulses that were stored and then retrieved from the memories (blue, $ V=(41.9\pm2.0)\%$) and non-stored ``leakage'' photons (gray, $V =(23.4\pm0.9)\%$). Error bars are statistical.}
  \label{fig:HOM_QM_pol}
\end{figure}

Histograms in Fig. \ref{fig:HOM_QM_pol}a,b show the storage of two $\rm \ket{D}$ polarized qubits from Alice and Bob in two dual-rail memories with a storage time of 1$\mu$s. The mean number of photons per pulse at the input of each memory is approximately $14$ and $10$ for Alice and Bob respectively. In the absence of the atomic ensemble, the transmission through each memory setup is measured to be approximately $3\%$ due to the low transmissions through each filtering etalon ($35\%$), the Faraday isolator ($50\%$), and various losses due to imperfect optics ($50\%$). Owing to slightly different storage and fiber coupling efficiencies, we measure average photon numbers per stored pulse of approximately $0.024$ and $0.017$ within a temporal region of interest (ROI) marked by the red and blue shaded regions in Fig. \ref{fig:HOM_QM_pol}a,b, from Alice and Bob's memory respectively. This ROI is post-selected with a width of $0.5\mu s$ beginning at the qubit retrieval time, following the procedure outlined in \cite{Noelleke2013}.\\

By varying the polarization of one qubit with respect to the other at the HOM interference node we observe the desired modulation in the coincidence rate, exhibiting a minimum for the initial identical polarizations and reaching a maximum corresponding to orthogonal qubit polarizations. The coincidences within the retrieval ROI, defined previously, are shown in red and fit (blue) to $\sim \cos^2$ in Fig. \ref{fig:HOM_QM_pol}c. The interference visibility is measured to be $V = (41.9 \pm 2.0) \%$ for the retrieved qubits, which is consistent with the measured visibility of $(42.1 \pm 0.2) \%$ without the memories (see Methods section). For the case of HOM interference without memories, we attribute the low visibility to polarization drifts in the long non-polarization maintaining fibers used. Thus, we conclude that our memories do not significantly affect the indistinguishability of the qubits. \\

The analysis detailed above is repeated with the two leftover ``leakage'' peaks, parts of the original input pulses that the memories do not store (gray shaded areas in Fig. \ref{fig:HOM_QM_pol}a,b). Here, a lower HOM visibility of $V=(23.4\pm0.9)\%$ is measured due to a mismatch in their temporal envelopes as well as photon numbers (data shown in gray with the dashed gray curve as the fit in Fig. \ref{fig:HOM_QM_pol}c).\\

\section{\large{III.  Gen I QM HOM interference at single-photon-level}} \label{sec:SinglePhotonHOMv1}
\begin{figure}[!ht]
\vspace{-2mm}
\centering
\includegraphics[width=1\columnwidth]{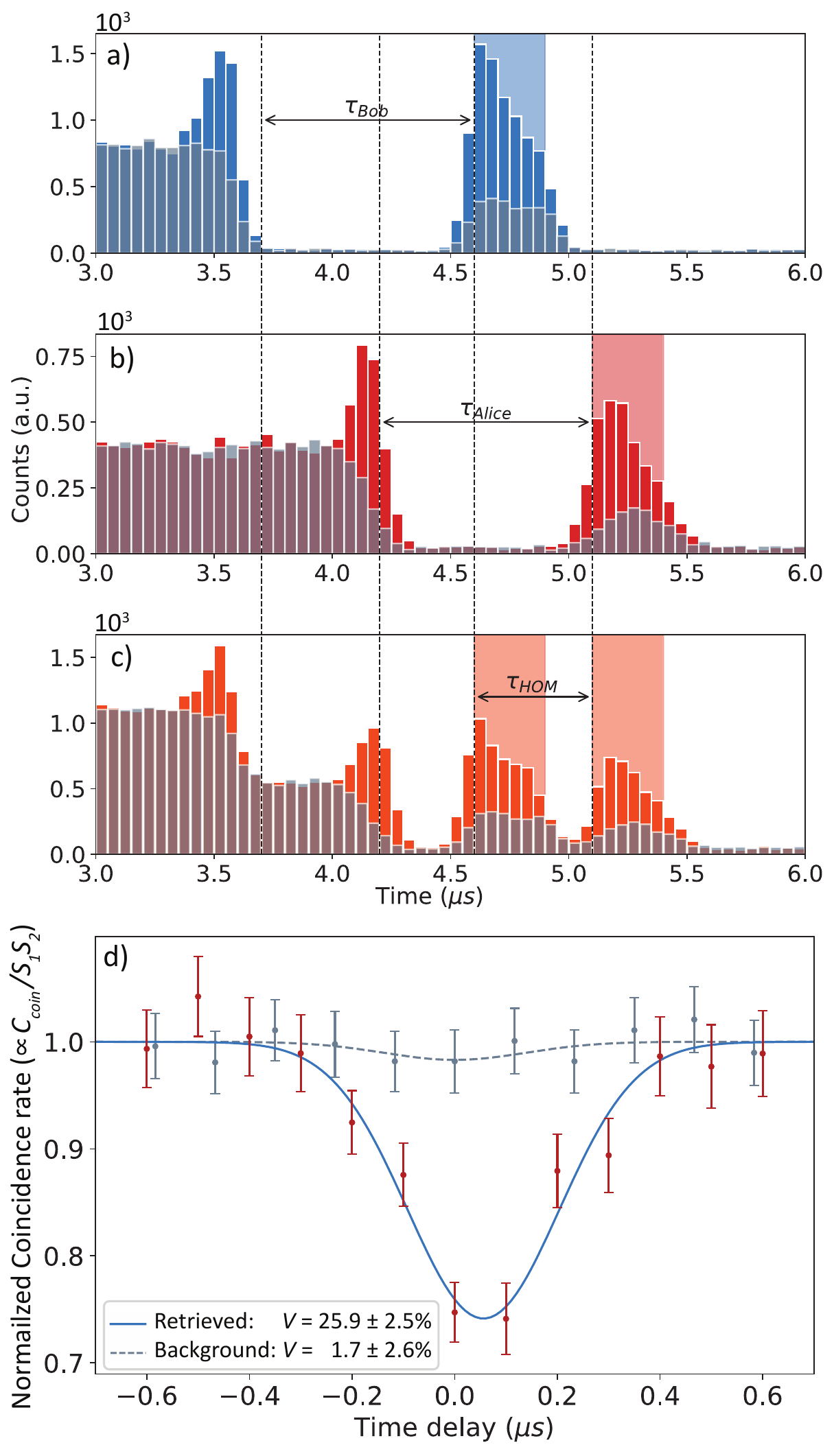}
\vspace{-2mm}
\caption{\footnotesize{{\bf HOM interference of pulses retrieved from two quantum memories with single-photon level inputs}. 
\textbf{(a)} Histograms of single-photon detection events for the storage of Bob's qubits containing 1.6 photons per pulse at the input of the memory; 
\textbf{(b)} same as \textbf{(a)} but for Alice's qubits containing 1.6 photons per pulse at the input of the memory. 
\textbf{(c)} Counts in the single photon counters when both Alice and Bob pulses are sent together. The storage times are equal, $\tau_{Alice} = \tau_{Bob} = 0.9 \mu s$. The time delay between retrieved pulses is indicated as $\tau_{HOM}$. The lightly shaded histograms in \textbf{(a), (b), (c)} show the received counts without the input probe pulses and only the control field pulses, indicating the amount of background noise. The vertical shaded areas indicate the region of interest.
\textbf{(d)} The normalized coincidence rate of retrieved pulses from memories at the measurement station (red points and blue fit). We obtain an interference visibility of $V = 25.9 \pm 2.5 \%$. The coincidence rate for background photons is also shown, not exhibiting the HOM dip (gray dots and line).}
}
\label{fig:HOM_1photon}
\end{figure}

To more closely mimic applications of these quantum memories in a network scenario and to investigate the effect of background, we reduce the input mean photon number of our qubits to contain on average approximately 1.6 photons per pulse. In this regime, the contribution of coincidences from background photons become significant. The memories are prepared in the same manner as the few-photon level experiment previously described, but this time the signal-to-background ratio (SBR) is measured to be around 2.6. Single photon level storage efficiencies of approximately $7\%$ and $18\%$ are measured for Alice's and Bob's memories, respectively, within the shaded red and blue temporal ROIs in Fig. \ref{fig:HOM_1photon}a,b. To account for this imbalance, the fiber coupling efficiencies are adjusted such that the mean photon numbers of the retrieved photons at the two inputs of the beamsplitter are matched.\\

Input pulses are chosen to be 400ns wide and the storage times for both the memories are the same ($\tau_{Alice} = \tau_{Bob} = 0.9\mu s$), however their arrival times, and consequently their retrieval times, are changed. The difference in retrieval times serves as the temporal delay for the HOM interference (Fig. \ref{fig:HOM_1photon}c). For each delay, in order to isolate the retrieved signal from the background, we looked for coincidences in the two detectors in the temporal region of interest (ROI) covering two $0.3\mu s$ regions in the retrieved pulses. Plotting the normalized coincidence rate versus the time delay, an HOM visibility of $(25.9\pm2.5)\%$ is obtained (Fig. \ref{fig:HOM_1photon}d), where the maximum obtainable visibility is measured to be $(42.4\pm0.6)\%$ (see Methods section). \\

Comparing results from sections II and III, we find that lowering the mean number of photons in the input lowers the SBR and subsequently lowers the visibility. To explore this effect, the experiment is repeated with only background and without the input qubits into the memory. As shown in gray dots in Fig. \ref{fig:HOM_1photon}d, an HOM visibility that is consistent with zero is found; demonstrating the noise sources do not exhibit second-degree interference.\\

\section{\large{IV.  Gen II QM HOM interference at single-photon-level}} \label{sec:SinglePhotonHOMv2}
\begin{figure}[!h]
  \centering
  \includegraphics[width=1\linewidth]{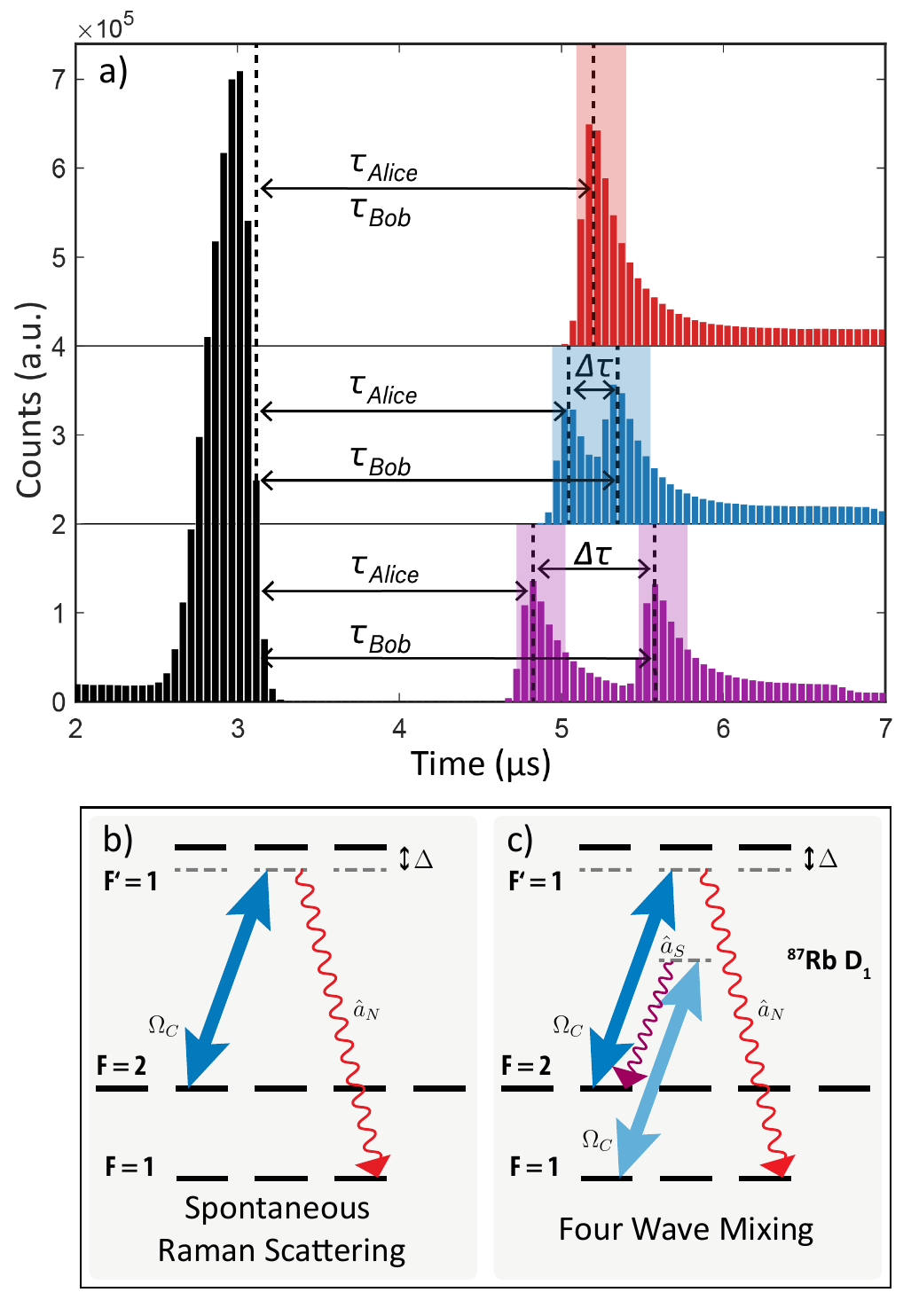}
  \caption{\footnotesize{\textbf{Varying storage times for HOM interference with high SBR quantum memories with single photon level inputs}. \textbf{(a)} Histograms of single-photon detection events from pulses, containing 1.6 photons per pulse, stored in two independent room temperature quantum memories. The difference in storage times, $\Delta \tau$, is used as the delay for HOM interference. (black: histogram of events showing a portion of the unstored pulse, red: histogram of simultaneous storage in Alice and Bob memories, each with a storage time of $\tau = 2.0\mu$s, blue: histogram of storage events with a difference in storage times of $\Delta\tau = 0.3\mu$s, purple: difference in storage times $\Delta\tau = 0.75 \mu$s). Shaded areas indicate an example of one region of interest used. \textbf{(b)}. The atomic levels and fields relevant for noise induced by Spontaneous Raman Scattering(SRS). SRS occurring due to the resonant control, shown in dark blue, results in an anti-Stoke's field, $\hat{a}_N$, (green) at the same frequency as the qubit, contributing to background noise. \textbf{(c)} The atomic levels and fields relevant for noise induced by Four Wave Mixing (FWM). The first step of FWM is the off-resonant coupling of the control, shown in light blue to the probe's transition, creating a Stokes field, $\hat{a}_S$, shown in purple, followed by the resonant control field creating an anti-Stokes field, $\hat{a}_N$ shown in green, at the same frequency and polarization as the qubit. Compared to our first generation quantum memories where the probe and control are linearly polarised, fewer channels are available for FWM and SRS to occur, reducing the amount of background in our second-generation quantum memories.}}
  \label{fig:procedureillustration}
\end{figure}

After experimental verification, three main sources of background in our memory setups are identified, all related to the control field employed to achieve the transparency window required for storage. In increasing order of significance, they are: the amplified spontaneous emission (ASE) coming from the tapered amplified laser used for the control field, the unfiltered control field, and the background due to atomic effects such as four-wave-mixing (FWM) and spontaneous Raman scattering (SRS) that produce noise with similar frequency and polarization to that of the qubits (Fig. \ref{fig:procedureillustration}b,c).\\

In our next set of experiments, these sources of background are reduced using various techniques that include a Volume Bragg Grating with a FWHM of 50GHz to remove the background associated with ASE, and new Fabry–Pérot etalons with a higher transmission for the probe field and higher suppression of the control. In addition to these changes, the utilization of $\sigma^\pm$ polarization for the probe and control fields suppresses\cite{Zhang2014} the background from FWM and SRS. All these improvements are implemented in our second-generation quantum memories that were jointly developed with Qunnect Inc. \cite{Qunnect2022} based on a Stony Brook University patent \cite{MemPatent2022}. For the experiments presented, we measure an SBR ranging from 1.7 to 11.9 depending on the size of the region of interest chosen, accompanied by an improved overall transmission of approximately $20\%$. \\


Gaussian pulses with a FWHM of $300$ns and a mean photon number per pulse of approximately $1.6$ serve as the inputs to the memories. After storage and retrieval from each memory, we measure the mean signal photon number to be approximately $0.01$ at the inputs of the beamsplitter, excluding the background photons for the entirety of the retrieved pulse.\\

In this experiment, we utilize the full potential of our quantum memories by taking advantage of the arbitrary storage time feature. Here, pulses created by Alice and Bob reach the memories at the same time and the HOM interference is observed by varying the difference in storage times of the memories, $\Delta \tau \equiv \tau_{Alice} - \tau_{Bob}$. The storage times are varied by equal and opposite amounts ($\Delta \tau/2$), with the sum of Alice's and Bob's storage times staying constant at $4 \mu s$ (Fig. \ref{fig:procedureillustration}a). For each run, $\Delta \tau$ is changed 25 times, and each time it is assigned a value at random from a predefined array in the range of -1.5$\mu$s to 1.5$\mu$s. Data is collected for 14 such runs for a total measurement time of $9$ hours.\\

\begin{figure}[!ht]
 \vspace{-2mm}
 \centering
 \includegraphics[width=1\columnwidth]{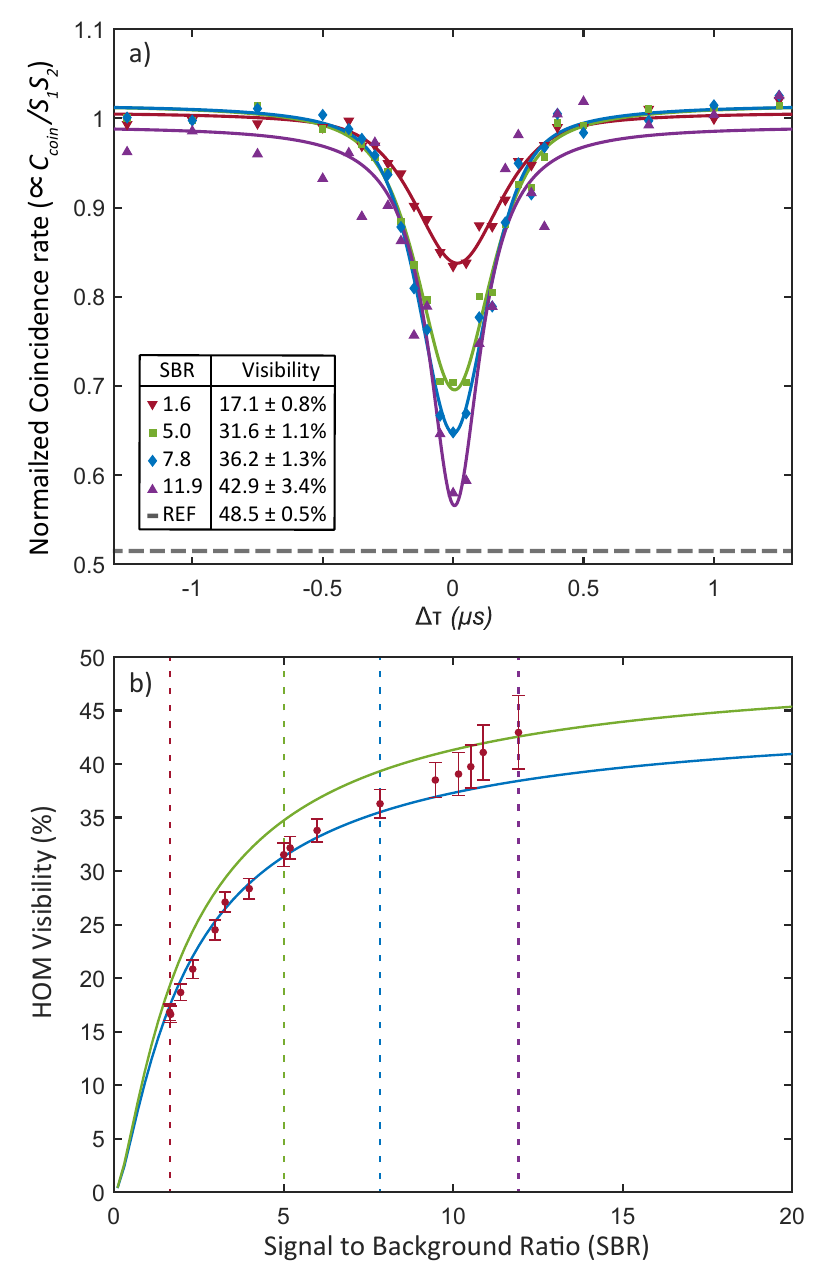} 
 \vspace{-2mm}

 \caption{\footnotesize{{\bf Dependence of HOM visibility on the signal-to-background ratio.}
 \textbf{(a)} HOM interference plotted for different estimated SBRs. Plots shown in red, green, blue, and purple show the HOM interference between photons that have been retrieved from the memories. These have visibilities of $17.1 \pm .8\%$, $31.6 \pm 1.1 \%$, $36.2 \pm 1.3 \%$, and $42.9 \pm 3.4 \%$ respectively. The black dashed line marks the measured visibility of $48.5 \pm .5 \%$ between pulses when no memories were involved. As the estimated SBR increases from 1.6 in the red plot to 11.9 for the purple plot, the HOM dip gets closer to the limiting case of no memories shown as the black dashed line. Error bars are not shown for clarity but can be found in Fig. \ref{fig:HOMDipsGenII}.
 \textbf{(b)} Measured HOM visibility, red dots, versus the SBR. All data points were analyzed from the same data set. Each SBR reported is estimated from the number of counts within the regions of interest used and a region of the same size of only background counts. The blue curve is obtained by a fit to Eq. \ref{eqn:VISvsSBR} where $V_0$ is the fit parameter, and reaches $45.5\%$ in the limit of infinite SBR. The green curve shows the ideal case of $V_0 = 0.5$. The dotted red, green, blue, and purple lines mark the SBRs plotted in \textbf{(a)}, respectively.}}
 \label{fig:visibilitymodel}
\end{figure}

For every $\Delta \tau$, coincidences between the two detectors within specific temporal ROIs are analyzed with a wide coincidence window to cover the entirety of the retrieved pulse. By changing the size of ROI, we significantly change the signal count rate that gets included, while keeping the background count rate relatively constant, hence changing the SBR. Analyzing different regions of interest we find a maximum SBR of approximately $11.9$, with a ROI of $160$ns. Using the data within this ROI, the normalized coincidence rate versus $\Delta \tau$ is plotted in Fig. \ref{fig:visibilitymodel}a; purple upward triangle. We measure an HOM visibility of $V=(42.9\pm3.4)\%$ with a new reference of $V=(48.5\pm0.5)\%$ measured without the memories (see Methods section). As before, the coincidence rate between the background noise photons exhibited no HOM interference.\\

To evaluate the significance of such an increase in visibility, a quantitative analysis of the effect of a non-interfering field on the HOM visibility is presented. In the absence of noise from the background, the retrieved signal pulses from the memories arrive simultaneously and with equal magnitude at the two input ports of a 50:50 beam splitter. Assuming identical detectors, these pulses are detected with a probability of $S_1=S_2 \equiv S$ per detector, and coincidences between signal photons are detected with a probability of $C_{coin}$. In this ideal scenario of identical detectors and no noise, visibility can be defined as

\begin{equation}
    V_0 \equiv 1-\frac{C_{coin}}{S^2}. \label{eqn:VISvsSBRIdeal}
\end{equation}

With the incorporation of background fields from the memories, additional coincidences can occur due to background-signal and background-background detections. Assuming that the signal and background mean photon numbers are low, one obtains a total coincidence rate of $C_{coin} + 2SB + B^2$ and a single detection rate of $S+B$, where $B$ is the background photon detection rate. Factoring in these effects, one obtains a simple expression demonstrating the detrimental effects of a non-interfering background on HOM visibility:

\begin{align}
    V &= 1-\frac{C_{coin} + 2SB + B^2}{(S+B)^2} \nonumber \\
    &= V_0 \left( \frac{R}{1+R}\right)^2, \label{eqn:VISvsSBR}
\end{align}

where $R$ is the SBR, and $V_0$ is the HOM visibility with no background. To show this dependence of HOM visibility on SBR, the above data analysis is repeated on the same data set for several ROIs, each resulting in a different estimated SBR (Fig. \ref{fig:visibilitymodel}b). A fit to the model (Eq. \ref{eqn:VISvsSBR}) indicates that the maximum achievable visibility is $45.5\%$, differing from the $48.5\%$ measured without the memories. We believe this is primarily due to a slight mismatch in memory efficiencies and slow drifts in polarization in the memory output fibers. Additionally, this model indicates we are nearing the plateau of visibility using quantum memories.

\section{\large{V.  Discussion}} \label{sec:Discussion}


\begin{table*}[!ht]
\begin{tabular}{| b{0.2mm} >{\centering}b{1.0in} >{\centering}b{0.3in} >{\centering}b{0.5in} >{\centering}b{0.6in} >{\centering}b{0.85in} >{\centering}b{0.4in} b{0.2mm} |}
\hline
& {  \bf Config. } & $\boldsymbol{\braket{n}}$ & 
{\bf DOF} & \bf{ROI}($\mu s$) & $\boldsymbol{V_{HOM}}$ & \bf{SBR} &\\
\cline{2-7}
& \multirow{3}{*}{\shortstack[l]{No memories}} & 0.4 & pol & N/A & $(42.1\pm0.2)\%^{*}$ & $\mathcal{O}(10^3)$ &\\ 
& & 0.4 & delay & N/A & $(42.4\pm0.6)\%^{*}$ & $\mathcal{O}(10^3)$ &\\ 
& & 0.5 & delay & N/A & $(48.5\pm0.5)\%^{\dagger}$ & $\mathcal{O}(10^3)$ &\\ 
\cline{2-7}
& \multirow{2}{*}{\shortstack[l]{Gen I Dual-rail}} & 13 & \multirow{2}{*}{\shortstack[l]{pol}} & 1.00 & $(35.8\pm1.7)\%^{*}$ & \multicolumn{1}{c}{24} &\\
& & 13 & & 0.50 & $(41.9\pm2.0)\%^{*}$ & \multicolumn{1}{c}{37} &\\
\cline{2-7}
& \multirow{2}{*}{\shortstack[l]{Gen I Single-rail}} & 1.6 & \multirow{2}{*}{\shortstack[l]{delay}} & 1.30 & $(20.3\pm2.3)\%^{*}$ & \multicolumn{1}{c}{2.4} &\\
& & 1.6 & & 0.60 & $(25.9\pm2.5)\%^{*}$ & \multicolumn{1}{c}{2.6} &\\
\cline{2-7}
& \multirow{3}{*}{\shortstack[l]{Gen II Single-rail}} & 1.6 & & 1.28 & $(31.6\pm1.1)\%^{\dagger}$ & \multicolumn{1}{c}{5.0} &\\
& & 1.6 & storage & 0.64 & $(36.2\pm1.3)\%^{\dagger}$ & \multicolumn{1}{c}{7.8} &\\
& & 1.6 & & 0.16 & $(42.9\pm3.4)\%^{\dagger}$ & \multicolumn{1}{c}{11.9} &\\
\cline{2-7}
& \multicolumn{6}{l}{$^{*}$Using the measurement station for the Gen I memories} &\\
& \multicolumn{6}{l}{$^{\dagger}$Using the measurement station for the Gen II memories} &\\
\cline{2-7}
\hline
\end{tabular}
\vspace{1mm}
\caption{\textbf{Measured HOM visibilities for Gen I and II experiments}: Different experimental configurations adopted (without memories, with dual rail memory operation, and with single rail memory operation). The first two reported results without memories were with a setup used in the experiments with the Gen I memories, and the third result without memories were with a setup used in the experiments with the Gen II memories. $\boldsymbol{\rm \langle n \rangle}$: mean photon number at the input of the memories or BS(in case of no memory operation), \textbf{DOF}: degree of freedom that is scanned to obtain the HOM dip. Here, Pol: polarization of Alice's qubit wrt Bob's, Delay: time of arrival of Alice's qubit wrt Bob's, Storage Time: the storage time of Alice wrt to Bob's. \textbf{ROI}: width of the region of intrest (ROI) time window to record coincidences, $\boldsymbol{V_{HOM}}$: measured visibility, \textbf{SBR}: signal-to-background ratio.}
\label{tab:mainresults}
\end{table*}

A comparison of results presented in section II and III highlight that the main parameter of interest is the SBR. We report a steady improvement in this and other memory parameters in section IV accompanied by a consistent increase in the measured visibilities as shown in Table \ref{tab:mainresults}. \\

These experiments present a clear road map to achieve high-visibility memory HOM interference of polarization qubits. In addition to the standard requirements for high visibility HOM such as spatiotemporal, frequency, and polarization matching between the interfering qubits, we also demonstrate the requirement of high SBR for high visibility. While background reduction techniques have been deployed in our systems, further improvements can be achieved by utilizing cavity mode selection, optical pumping, and active manipulation of the EIT Hamiltonian. Furthermore, it is possible to increase  the concentration of the signal with respect to the background by using a shorter qubit pulse on the condition of no adverse effect on efficiency \cite{Wolters2017}. Short pulses also have the added benefit of more closely mimicking the temporal profile of photons retrieved from cavity-enhanced SPDC entanglement sources such as the ones we deploy in our laboratory. We predict that narrowing pulse widths by a factor of 10, i.e. $300$ns $\rightarrow 30ns$, will likely result in an order of magnitude increase in the SBR reaching $\sim$120. In conjunction with the use of a superconducting nanowire detector to effectively decrease our dark counts to zero, we believe we will be able to achieve a HOM visibility of nearly $50\%$ with our second-generation QMs.\\

\section{\large{VI.  Outlook}} \label{sec:Outlook}

\subsection{A.  Towards a type II quantum repeater}
In terms of the scalability required to implement large-scale systems, we have implemented and demonstrated the interoperability of four room-temperature atomic quantum memories, the minimum number of memories required for a type II quantum repeater. We envision using our second-generation QMs to facilitate entanglement swapping at the central node, while in parallel, upgraded first-generation QMs buffer the external nodes. To implement the basic framework of a first-generation quantum repeater, high SBR storage of single-photons from a bright entanglement source will be crucial. Recent developments have demonstrated the storage of single photons originating from SPDC sources \cite{Buser2022}. Additionally, there have been successful interactions between single photons emitted by quantum dot sources and room-temperature atomic quantum memories \cite{Guodong2023, Kroh2019}. These cutting-edge techniques for photon pair generation represent the most promising contenders for a high repetition source of entangled photons with linewidths compatible with our atomic quantum memories.\\

Given the current advances in memory-source interactions with high repetition rate sources, heralding these interactions in real-time is the only major scientific milestone that needs to be met in order to realize a practical type II quantum repeater network. This milestone stands as the core focus of our forthcoming experiments which will build upon our previous work \cite{HeraldingSteve}, wherein we successfully detect the presence of a signal field in an atomic memory through a non-demolition measurement that employs quantum state tomography to measure the change in the phase of a probe field that co-propagates with the signal. We envision adopting this phase-phase photon-photon non-linear system to herald the presence of a single photon in our quantum memories in real-time.\\

\subsection{B.  Towards Memory Assisted MDI-QKD}
We anticipate that one of the several immediate applications of our type II repeater network topology could be buffering streams of random polarization qubits to be employed in a variable-delay MDI-QKD protocol \cite{Kaneda2017, Xu2013} and Bell-state measurements. The memory-assisted protocols require that Alice and Bob create qubits and send them through their respective channels where they are stored in their respective memories. Once both memories are loaded and heralded, they are simultaneously read, and the rest of the protocol is identical to standard MDI-QKD. We have adopted the approach from \cite{Panayi2014} to predict the secret key rate for a possible implementation of MA-MDI-QKD protocol using our network. We included a source of background in the analysis to study the dependence of key rate on the signal to background ratio of the signal retrieved from our memories. The model predicts that with our current Gen II memory parameters and an achievable signal to background of 135, it is possible to achieve a positive key rate. \\

The protocol requires that Alice and Bob create attenuated coherent states and send them through channels A and B. Alice's and Bob's memory each store the state sent by Alice/Bob. Once both memories are loaded, they are simultaneously read, and the rest of the protocol is identical to the standard MDI-QKD procedure. The quantities can be first derived for the original MDI QKD protocol and then modified to include memory effects.

In order to calculate the secret key rate for MA-MDI-QKD set up, we need the following quantities: $R_s$, $Q_{11}^{QM}$, $e_{11;X}^{QM}$, $Q_{\mu\nu; Z}^{QM}$ and $E_{\mu\nu;Z}^{QM}$, as defined below.\\

\begin{enumerate}
 \item $R_{s}$: the repetition rate at which weak coherent pulses are prepared by Alice and Bob.
 \item $Q_{\mu\nu; Z}^{QM}$ : probability that memories in both the channels are loaded with coherent states prepared in the rectilinear(Z) basis ($\ket{\mu}$ in Alice, $\ket{\nu}$ in Bob) and that a successful Bell State Measurement(BSM) is performed. 
 \item $Q_{11}^{QM}$ : contribution of single photon states to the term $Q_{\mu\nu; Z}^{QM}$.
 \item $e_{11; x}^{QM}$ : error rate for the case of single photon fock states in the diagonal basis and $E_{\mu\nu; Z}^{QM}$ is the error rate for coherent states in the rectilinear basis. 
 \item{f}: ratio between the actual cost of error correction to its minimum value obtained by the Shannon's Theorem. f is taken to be 1.16.
\end{enumerate}

The expressions for the above quantities have been taken from \cite{Panayi2014}, along with the expression for the secret key rate, given by,

\begin{widetext}
\begin{equation}
    R_{MA-MDI-QKD} = R_s \left[ Q_{11}^{QM} \left( 1-h(e_{11;X}^{QM}) \right) - f Q_{\mu\nu; Z}^{QM}h(E_{\mu\nu;Z}^{QM}) \right]
\label{eqn:R_MA-MDI-QKD}
\end{equation}
\end{widetext}

where $h(x)$ is the binary entropy function. 

\begin{table*}[!ht]
\begin{tabular}{| b{0.2mm} b{0.4in} b{0.4in} b{0.4in} b{0.5in} b{0.8in} b{0.5in} b{0.9in} b{0.2mm} |}
\hline
& \textbf{$\mu $} & T($\%$) & \textbf{$\eta_{w}$} & \textbf{$\eta_{r0}$} & \textbf{Rep rate} & \textbf{Min SBR} & \textbf{Key Rate at SBR of 200} & \\
\cline{2-8} 
& 1.6 & 16$\%$ & 1 & 0.05 & 200 kHz & 135 & $\mathcal{O}(10^{-1})$ & \\
\cline{2-8}
& 1 & 16$\%$ & 1 & 0.05 & 200 kHz & 55 & $\mathcal{O}(10^{-1})$ &  \\
\cline{2-8}
& 0.6 & 16$\%$ & 1 & 0.05 & 200 kHz & 50 & $\mathcal{O}(10^{-1})$ &  \\
\cline{2-8}
& 1.6 & 70$\%$ & 1 & 0.7 & 2 MHz & 135 & $\mathcal{O}(10^3)$ &  \\
\cline{2-8}
\hline
\end{tabular}
\vspace{1mm}
\caption{\textbf{Required SBR to achieve a positive key rate for different experimental parameters.} $\mu$ is the mean photon number of the input pulses, T is the overall optical transmission through the system, $\eta_{w}$ and $\eta_{r}$ are the writing and reading efficiencies of the system, with $\eta_{w}$ assumed to be 1. The last column in the table is the key rate for a signal to background ratio of 200.}
\label{tab:keyrate}
\end{table*}

The main source of background in our memories is from four wave mixing and spontaneous Raman scattering that creates a background at the same frequency as the signal. Hence the mean photon number of the background can be assumed to be $\mu/R$, where $\mu$ is the mean photon number of the signal and R refers to the signal to background ratio. This can be used to calculate the probabilities that the memories get loaded by background photons. Incorporating this source of background into the error rates, allows us to write the secret key rate as a function of the signal to background ratio, R. We see that it is possible to achieve a positive key rate if the threshold for signal to background is met. Table \ref{tab:keyrate} lists the minimum signal to background ratios required to achieve a positive key rate for different system parameters. We find that even with current system parameters it is possible to achieve a positive key rate, provided we can attain higher signal to background ratio. On improving parameters such as the overall transmission of the system, the reading efficiency as well as the repetition rate, we see that the achievable key rate increases by 4 orders of magnitude, compare line 1 and 4 in Table \ref{tab:keyrate}.

\section{\large{VII.  Methods}} \label{sec:Characterization}

\subsection{A.  Polarization qubit sources} 

\begin{figure*}[!ht]
  \vspace{-1mm}
  \centering
  \includegraphics[width=1\linewidth]{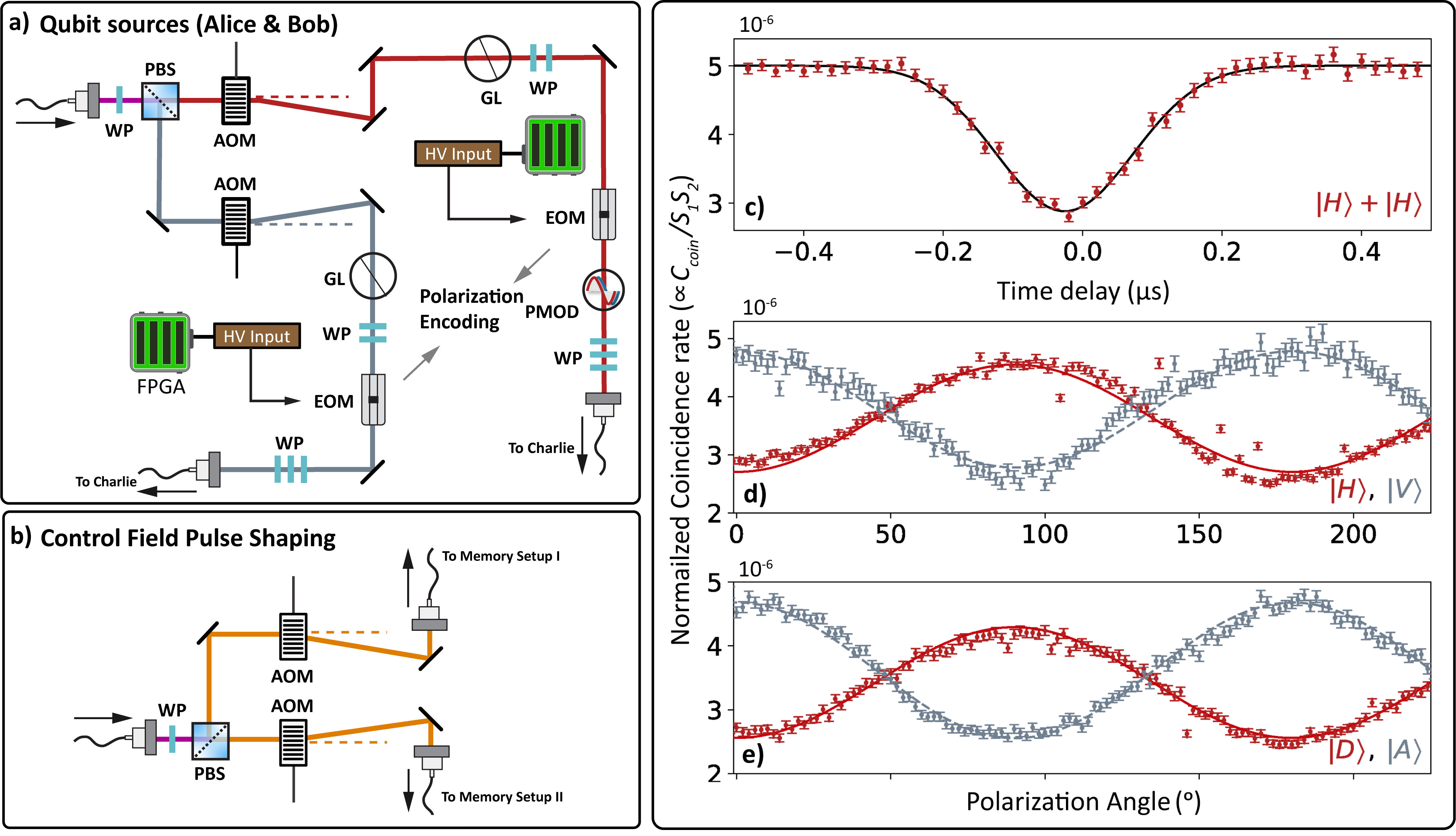}
  \caption{\textbf{HOM interference of two independent polarization qubit sources}. \textbf{(a)} Alice and Bob generate polarization qubit pulses using AOMs (Acousto-Optic Modulators, for pulse shaping) and EOMs (Electro-Optic polarization Modulators, for polarization qubit encoding). The phases of Bob's pulses are randomized by the phase modulator (PMOD). The prepared qubit streams can be routed directly to the measurement station, Charlie, or to their respective memories. \textbf{(b)} Pulse shaping of the control fields for storage and retrieval of qubits in the quantum memories. The control field for each memory setup is turned on for a few $\mu$s to prepare the atoms. For storage, the fields are turned off as the qubits propagate through the memories, and turned back on after an arbitrary time for retrieval. \textbf{(c)} HOM interference is measured without the memories, with qubits having a relative temporal delay by changing the pulse generation time of Alice wrt Bob; \textbf{(d), (e)} Interference experiments without memories are repeated for a fixed polarization of Alice's qubits while varying the polarization of Bob's qubits. Alice's qubits are fixed as $\rm \ket{H}$, $\ket{V}$ (shown in \textbf{(b)}), $\rm \ket{D}$ and $\ket{A}$ (shown in \textbf{(c)}). The interference visibilities are measured as $\rm V = (42.4\pm0.6)\%$ for the delay HOM experiment and $\rm V = (42.1\pm0.2)\%$ for the polarization HOM experiments. Error bars are statistical.\\
  \textbf{GL} Glan Laser Polarizer, \textbf{PBS} Polarising Beam Splitter.}
  \label{fig:HOM_charlie}
\end{figure*}

For each experiment presented, qubits are created by the sources named Alice and Bob. Each qubit source is sent attenuated coherent states from a 795nm diode laser. The laser is red detuned by 400MHz from the $ 1 \rightarrow 1^{'}$ transition in the D1 line of $^{87}Rb$. The level of attenuation is adjusted depending on the required photon number per pulse. As seen in Fig. \ref{fig:HOM_charlie}, the setup for both Alice and Bob consists of an acousto-optic modulator (AOM) and an electro-optic modulator (EOM). The AOMs are used to create pulses with a Gaussian temporal profile with varying FWHM for different experiments. Electro-optic-modulators are used for polarization encoding at frequencies of up to $\rm 1MHz$. To avoid first-order interference effects, a phase modulator (PM) is used to randomize the phase of Alice with respect to Bob. The phase is driven with triangle modulation at $\rm \sim 1kHz$ with an amplitude of $\pi$, to ensure the uniform sampling of all relative phases during a measurement. For all the measurements presented in this paper, the Gaussian pulses have a temporal FWHM ranging from 200-400ns, within the 1$\mu$s coherence time of our laser.\\

\subsection{B.  Quantum memories}

\begin{table*}[!ht]
\begin{tabular}{l|cc|cc|cc|}
\cline{2-7}
 &
  \multicolumn{2}{c|}{\begin{tabular}[c]{@{}c@{}}Gen I\\ Few Photon Level\end{tabular}} &
  \multicolumn{2}{c|}{\begin{tabular}[c]{@{}c@{}}Gen I\\ Single Photon Level\end{tabular}} &
  \multicolumn{2}{c|}{\begin{tabular}[c]{@{}c@{}}Gen II\\ Single Photon Level\end{tabular}} \\ \hline
\multicolumn{1}{|l|}{Memory}                       & \multicolumn{1}{c|}{Alice} & Bob & \multicolumn{1}{c|}{Alice} & Bob  & \multicolumn{1}{c|}{Alice} & Bob \\ \hline
\multicolumn{1}{|l|}{Mean Input Photons/Pulse}     & \multicolumn{1}{c|}{14}    & 10  & \multicolumn{1}{c|}{1.6}   & 1.6  & \multicolumn{1}{c|}{1.6}   & 1.6 \\ \hline
\multicolumn{1}{|l|}{Storage Efficiency}           & \multicolumn{2}{c|}{20\%}        & \multicolumn{1}{c|}{7\%}   & 18\% & \multicolumn{2}{c|}{5\%}         \\ \hline
\multicolumn{1}{|l|}{Filtering Transmission}       & \multicolumn{2}{c|}{3\%}         & \multicolumn{2}{c|}{3\%}          & \multicolumn{2}{c|}{17\%}        \\ \hline
\multicolumn{1}{|l|}{\begin{tabular}[c]{@{}l@{}}Memory Out\\ Coupling Efficiency\end{tabular}} &
  \multicolumn{2}{c|}{70\%} &
  \multicolumn{1}{c|}{48\%} &
  19\% &
  \multicolumn{2}{c|}{75\%} \\ \hline
\multicolumn{1}{|l|}{\begin{tabular}[c]{@{}l@{}}Mean Photon Number/Pulse\\ Incident on the Beam Splitter\end{tabular}} &
  \multicolumn{1}{c|}{0.059} &
  0.042 &
  \multicolumn{1}{c|}{0.0016} &
  0.0016 &
  \multicolumn{1}{c|}{0.0102} &
  0.0102 \\ \hline
\multicolumn{1}{|l|}{Detector Coupling Efficiency} & \multicolumn{2}{c|}{70\%}        & \multicolumn{2}{c|}{70\%}         & \multicolumn{2}{c|}{70\%}        \\ \hline
\multicolumn{1}{|l|}{Detector Efficiency}          & \multicolumn{2}{c|}{58\%}        & \multicolumn{2}{c|}{58\%}         & \multicolumn{2}{c|}{58\%}        \\ \hline
\multicolumn{1}{|l|}{\begin{tabular}[c]{@{}l@{}}Retrieved Signal \\ Photons/Pulse Detected\end{tabular}} &
  \multicolumn{1}{c|}{0.0239} &
  0.0170 &
  \multicolumn{1}{c|}{0.00065} &
  0.00065 &
  \multicolumn{1}{c|}{0.0041} &
  0.0041 \\ \hline
\end{tabular}

\vspace{1mm}
\caption{\textbf{A Direct Comparison of Various Efficiencies Between Gen I and Gen II memories.} The reported values for the Generation I memories were counted in the region of interest used for their analysis, which was defined to be $.6 \mu s$ with an SBR of 2.6. The reported values for the Generation II memories were counted using the majority of the retrieval pulse. Specifically, within a region of $.64 \mu s$ defined by $35 \% $ of the peak height with an SBR of 7.9.}
\label{tab:Gen1&2}
\end{table*}
We present results from two generations of quantum memories, Fig. \ref{fig:MemoryDiagram}a,b, which are described below. Table \ref{tab:Gen1&2} provides a direct comparison of memory efficiencies under different conditions and across the generations. The storage efficiencies reported are calculated using the retrieved photons within a ROI $\sim 0.6 \mu s$. Both generations of the memory operate via Electromagnetically Induced Transparency (EIT) realized in a warm atomic vapor in a $\rm \Lambda$ configuration within the D$_1$ line of $\rm ^{87}Rb$, with a probe field tuned to the $1 \rightarrow 1^{'}$ transition and a control field coupled to the $2 \rightarrow 1^{'}$ transition as seen in Fig. \ref{fig:MemoryDiagram}d. The probe field is red-detuned from resonance by 400MHz, an optimal detuning for our memories as estimated in \cite{Namazi2017}. The control field is phase-locked to an offset of $\rm 6.835GHz$ with respect to the probe laser, thus imposing a zero two-photon detuning condition between the probe and control fields. Storage of polarization qubits is achieved in a dual-rail configuration, where beam displacement optics (BDO) placed at the input of the memory map an arbitrary superposition of $\ket{H}$ and $\ket{V}$ polarization modes onto a spatial superposition of the left and right rails. The two rails of the control beam coherently prepare two volumes in the $\rm ^{87}$Rb vapor cell, enabling quantum storage in both rails. The stored qubits can be retrieved on-demand by changing the timing of the ``read'' control pulse. \\

The storage time is controlled by varying the time for which the control field stays off. This is achieved by appropriately modulating the amplitude of the control field. After successfully retrieving the stored qubits, polarization elements are employed to obtain $\rm 42dB$ of control field extinction. Subsequently, frequency filtering setups, consisting of two consecutive Fabry-Perot (FP) cavities - each providing at least $\rm 50dB$ of control field extinction, are used to further remove the control field photons. The cascaded etalons are chosen to have a free spectral range (FSR) such that it minimizes the transmission of the control field, which is locked at an offset 6.834GHz from the probe field. The etalons have a free spectral range (FSR) of $13$GHz and $21$GHz, to provide $>80$dB control field suppression and are separated by a Faraday Isolator (FI), modified to be insensitive to the polarization mode. Finally, the outputs of the filtering systems are fiber-coupled and sent to the measurement node, Charlie.\\

\subsubsection{1.  Generation I, Table Top Memories} 
The table top memory set-ups consist of isotopically pure $\rm ^{87}Rb$ atoms in a $\rm 65 mm$ long vacuum-sealed glass cell maintained at a temperature of $\rm \sim 60^{\circ}C$. The cells also contain a $\rm Kr-Ne$ buffer gas to reduce decoherence. Each atomic vapor cell is placed at the center of three concentric \textit{Mu}-metal cylindrical shields creating a magnetic-field-free environment. The Lorentzian bandwidths of the etalons were $\sim$ 22 MHz and $\sim$ 17 MHz for Alice's memory system, and $\sim$ 15 MHz and $\sim$ 14 MHz for Bob's memory system.\\

\subsubsection{2.  Generation II, Portable Memories} 
Similar to the previous generation, the light-matter interface of the rack-mountable memory set-ups also consists of isotopically pure $\rm ^{87}Rb$ atoms with a $\rm Ne$ buffer gas in a $\rm 80 mm$ long vacuum-sealed glass cell maintained at a temperature of $\rm \sim 50^{\circ}$. The vapor cells are placed at the center of two concentric \textit{Mu}-metal cylindrical shields creating a near magnetic field-free environment. The primary difference between the Gen II and Gen I memories is the use of $\sigma^\pm$ polarization for the probe and control fields, which allows for fewer channels for the creation of four-wave mixing, hence decreasing the amount of background noise. In addition, we obtain improved control field filtering, and improved overall transmission. More information about these memories can be found in \cite{Qunnect2022}.\\

\subsection{C.  HOM Measurement Station Calibration}
For all delay experiments, with and without the memories, a series of wave plates are employed at the Charlie station before the beam splitter in each arm to correct for any unitary transformations of the polarization during propagation through the fibers. This transformation is analyzed periodically, outside of measurement time, using a continuous wave reference through the entire setup and measured with a polarimeter. The pulses retrieved from the two room-temperature quantum memories interfere at a $\rm 50:50$ beamsplitter, and two single-photon detectors (SPCMs) placed at the output arms of the beamsplitter generate a signal every time they record a hit. Data from the SPCMs is analyzed to calculate the coincidence rate between the two output arms of the NPBS.\\

To calibrate the measurement station for different polarization states, the qubits are routed directly to the measurement station, bypassing the memories. By varying Bob's qubit's polarization with respect to Alice's, the relative polarization between the qubits is utilized to measure HOM interference. In this case, 400ns Gaussian pulses containing $\approx 0.4$ photons per pulse are used. Eventually, as seen in Fig. \ref{fig:HOM_charlie}d,e, perpendicular qubit polarizations result in two entirely distinguishable photons with a maximum coincidence rate at the outputs of the beamsplitter. The normalized coincidences are fitted to $cos^{2}(\phi)$, where $\phi$ is the relative polarization.  To ensure that the setup is balanced for all the input states, polarization HOM interference is observed for all four-qubit states of Alice, showing on average a visibility of $(42.1\pm$0.2)\%. This measurement set-up is deployed for measurements presented in section IV.\\

To more closely mimic the delay experiments with the memories, the time delay, thus the temporal overlap of the qubit pulses are scanned with the qubits once again routed around the memory. Both Alice and Bob create 200ns wide Gaussian pulses that are horizontally polarized and contain on average $\approx 0.4$ photons per pulse. The temporal overlap of the two wave functions is scanned by changing the timing of pulse generation between Alice and Bob. The rate of coincidence counts between the output ports of the beamsplitter is measured using a coincidence window of $1 \mu s$. We measure a visibility of $V = (42.4\pm0.6)\%$ from the fit at zero delay, shown in Fig. \ref{fig:HOM_charlie}c. As expected \cite{Legero2004}, the width of the HOM curve precisely matches the temporal width of the input pulses. For the experiments with the Gen II memories, presented in section IV, we use a new measurement station with a visibility of at least $V = (48.5 \pm 0.5)\%$ (Fig. \ref{fig:HOMDipsGenII}f). These measured visibility values serve as a reference for the HOM interference with qubits retrieved from the quantum memories.\\

\subsection{D.  Automation and control}
\begin{figure*}[!ht]
 \centering
 \includegraphics[width=1\linewidth]{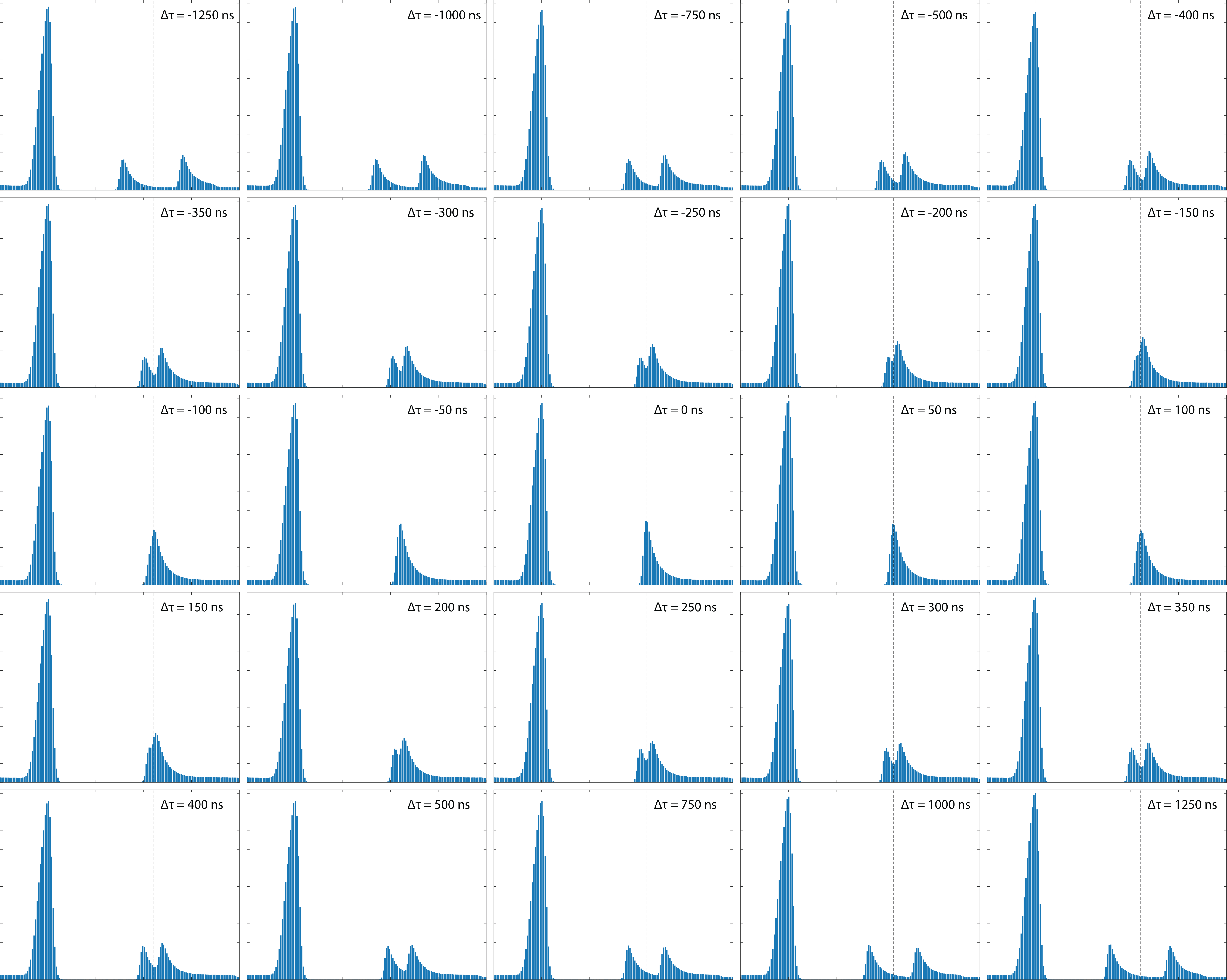}

 \caption{Histograms of single-photon detection events showing the unstored and retrieval pulses where the input, $\braket{n}\approx 1.6$, was stored in two independent room temperature quantum memories for every $\Delta \tau$ used. The X-axis of each plot has units of $\mu s$ and each Y-axis has arbitrary units as it displays the histogram count.}
 \label{fig:Procedure}
\end{figure*}

For the first set of single-photon level experiments (section III), the data was taken for approximately $\rm 20$ hours at a repetition rate of $\rm 40kHz$ with the time of arrival of pulses to one memory scanned with respect to another. To obtain near-identical conditions for each time delay segment, each time delay scan takes 120 seconds. For the second set of single-photon level experiments (section IV), the data was taken for approximately $\rm 9$ hours at a repetition rate of $\rm 100 kHz$ with the storage times of both memories scanned. Each difference in storage time, $\Delta \tau$, was held for either $1$ or $2$ minutes before being changed. A cumulative histogram of the storage and retrieval from both detectors and memories is shown in Fig. \ref{fig:Procedure} for every $\Delta \tau$ analyzed. For each set of experiments, conditions were periodically optimized between runs to guarantee that the transmission through the filtering system was relatively stable throughout the data-taking procedure.\\

\begin{table}
\begin{tabular}{| b{0.2mm} b{0.45in} b{0.45in} b{0.45in} b{0.45in} b{0.45in} b{0.2mm} |}
\hline
& {\bf ROI ($\mu s$)} & $\boldsymbol{\braket{n_{tot}}}$ & $\boldsymbol{\braket{n_{sig}}}$ & {\bf SBR} & $\boldsymbol{r_{coin}}$ (Hz) &\\
\cline{2-6}
& 4.300 & 0.0175 & 0.0109 &  1.67 & 7.18 &\\  
& 4.023 & 0.0173 & 0.0107 &  1.64 & 7.09 &\\  
& 3.400 & 0.0160 & 0.0106 &  1.96 & 6.57 &\\  
& 2.800 & 0.0150 & 0.0105 &  2.31 & 5.99 &\\  
& 2.200 & 0.0140 & 0.0105 &  2.98 & 5.36 &\\  
& 1.900 & 0.0135 & 0.0103 &  3.27 & 4.98 &\\  
& 1.686 & 0.0134 & 0.0107 &  3.98 & 4.65 &\\  
& 1.279 & 0.0124 & 0.0103 &  5.00 & 3.93 &\\  
& 1.208 & 0.0122 & 0.0102 &  5.19 & 3.79 &\\  
& 0.990 & 0.0114 & 0.0098 &  6.00 & 3.30 &\\  
& 0.638 & 0.0093 & 0.0082 &  7.84 & 2.30 &\\  
& 0.453 & 0.0076 & 0.0069 &  9.47 & 1.63 &\\  
& 0.322 & 0.0060 & 0.0055 & 10.16 & 1.06 &\\  
& 0.290 & 0.0055 & 0.0050 & 10.52 & 0.91 &\\  
& 0.240 & 0.0048 & 0.0044 & 10.89 & 0.69 &\\  
& 0.157 & 0.0032 & 0.0030 & 11.93 & 0.35 &\\  
\hline
\end{tabular}
\vspace{1mm}
\caption{\textbf{Measured Photon Statistics per Pulse for Gen II Memories for all Regions Analyzed.} For the analysis involving the second generation of memories various ROI widths were used to obtain different signal-to-background ratios (SBR). Here, the measured average mean photon numbers per pulse are reported across all $\Delta \tau$ used. $\braket{n_{tot}}$ is the total average photon numbers within the reported ROI for both memories per pulse, including the signal and background counts. $\braket{n_{sig}}$ reports the estimated number of signal photons collected within the reported ROI for both memories per pulse, this was calculated by subtracting the number of estimated background counts from the total number of counts. The number of background counts was estimated by looking at a similarly sized ROI that was known to contain only background and dark-count photons. $r_{coin}$ gives the average coincidence rate when the retrieval photons are distinguishable ($\Delta \tau = \pm 1250ns, \pm 1000ns$) under a repetition rate of $100$kHz.}
\label{tab:ROI}
\end{table}

\subsection{E.  Analysis}
In each interference experiment, two fiber-coupled single-photon counting modules (SPCM) detect the photon events with an efficiency of approximately $58\%$ at each branch of the beamsplitter, and a time-tagger unit registers the photon arrival time with an accuracy of $\rm 100ps$ and an overall jitter of $\rm 10ns$. \\

\begin{figure*}[!ht]
 \includegraphics[width=0.32\linewidth]{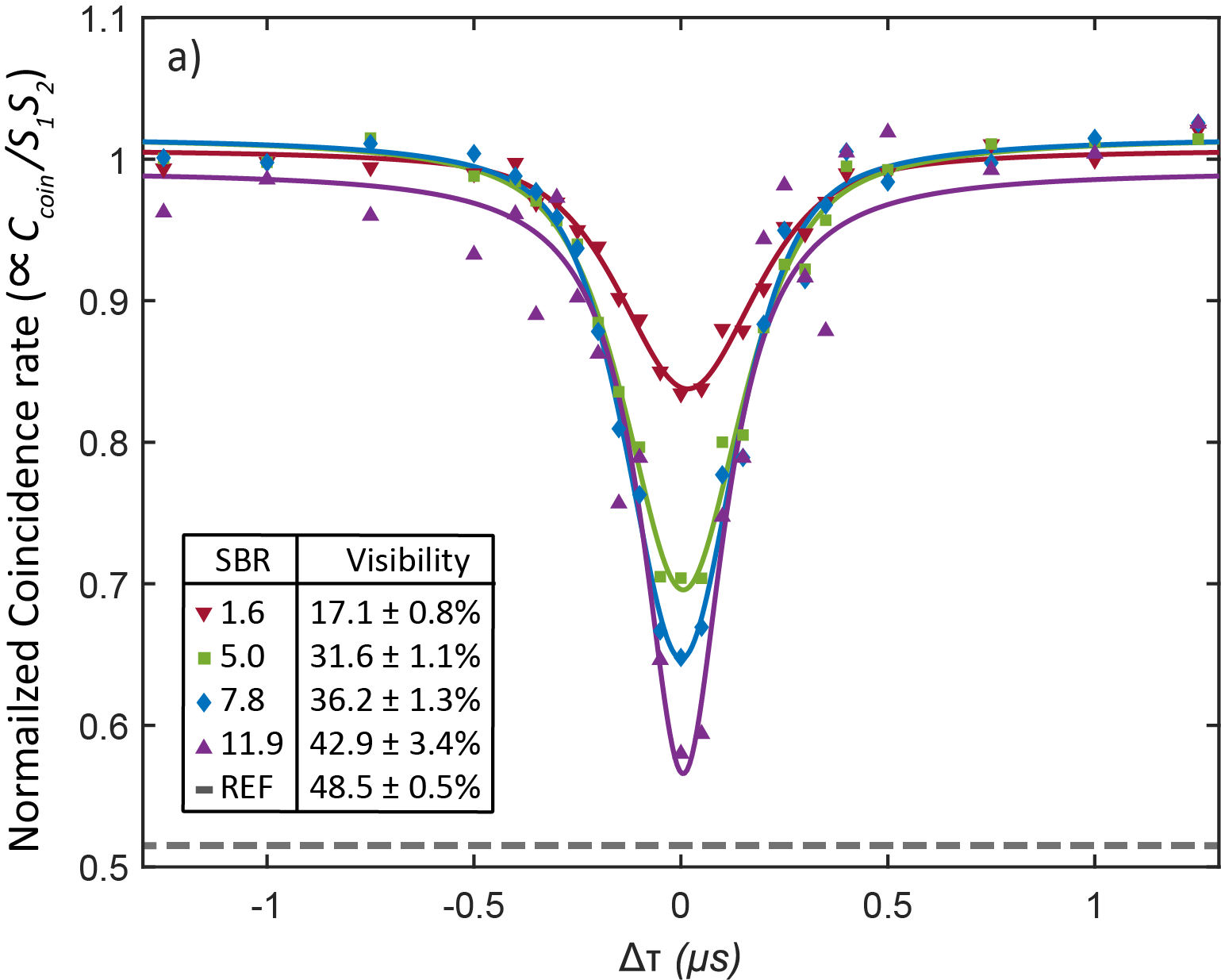}
 \includegraphics[width=0.32\linewidth]{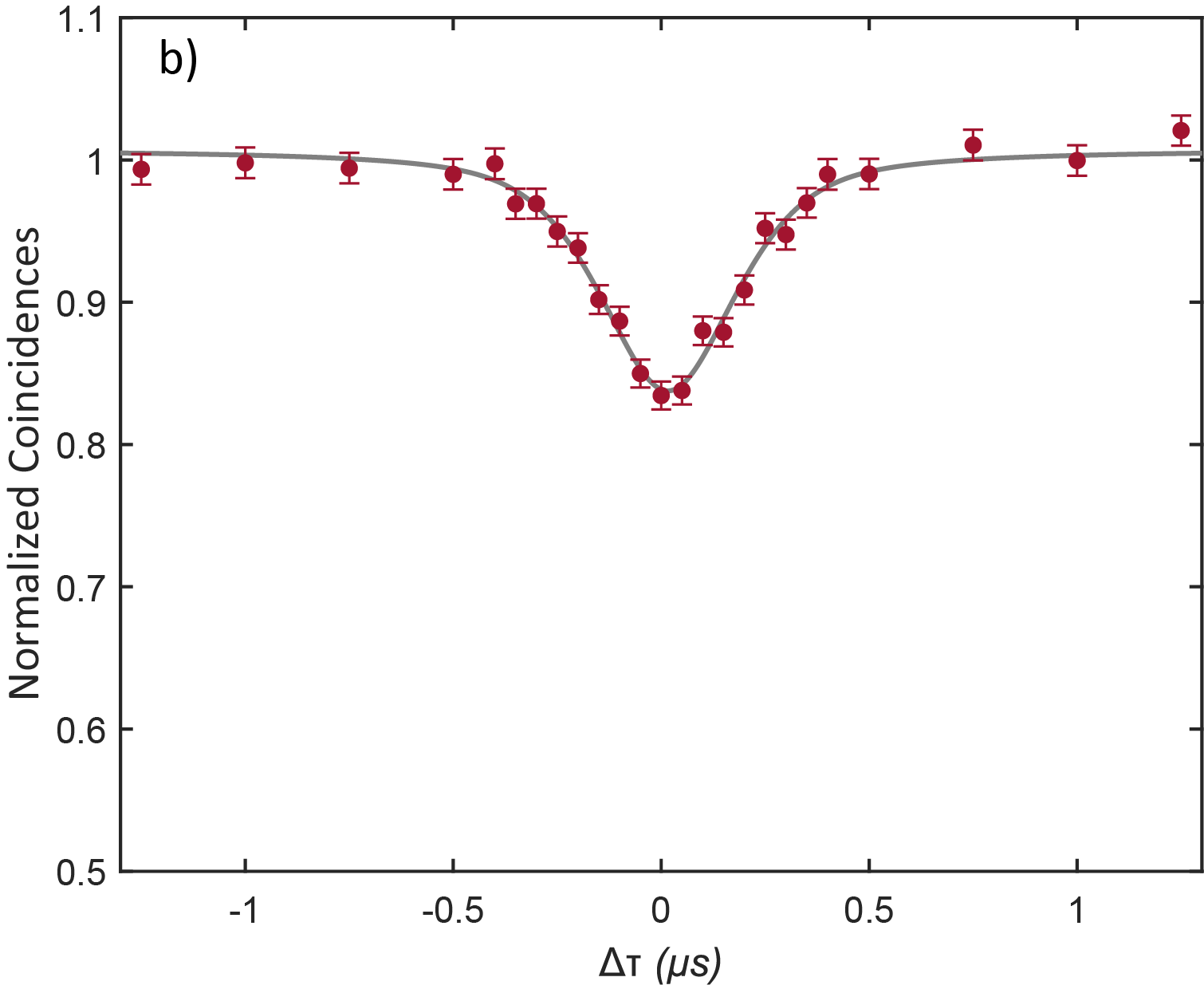}
 \includegraphics[width=0.32\linewidth]{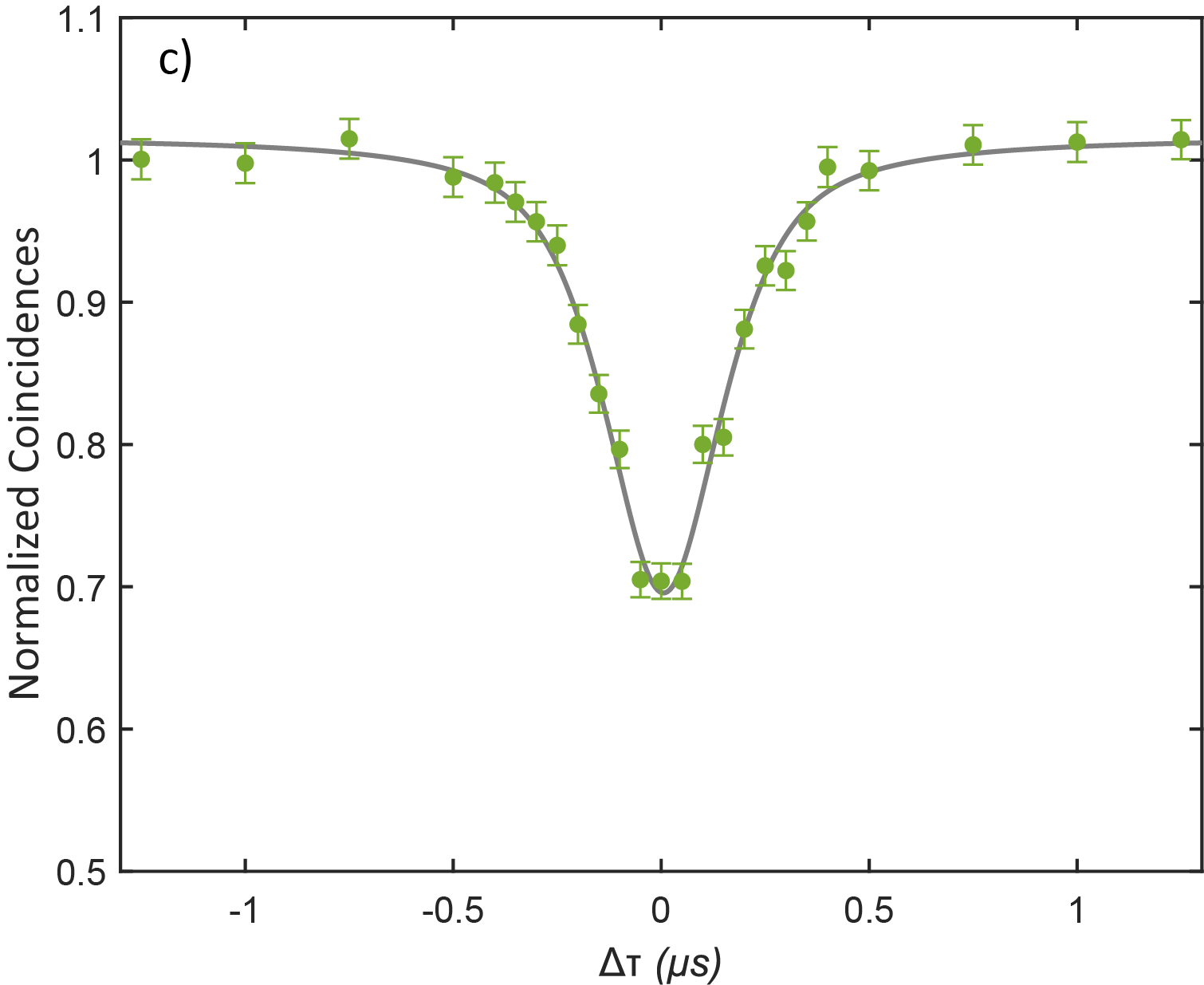}
 \includegraphics[width=0.32\linewidth]{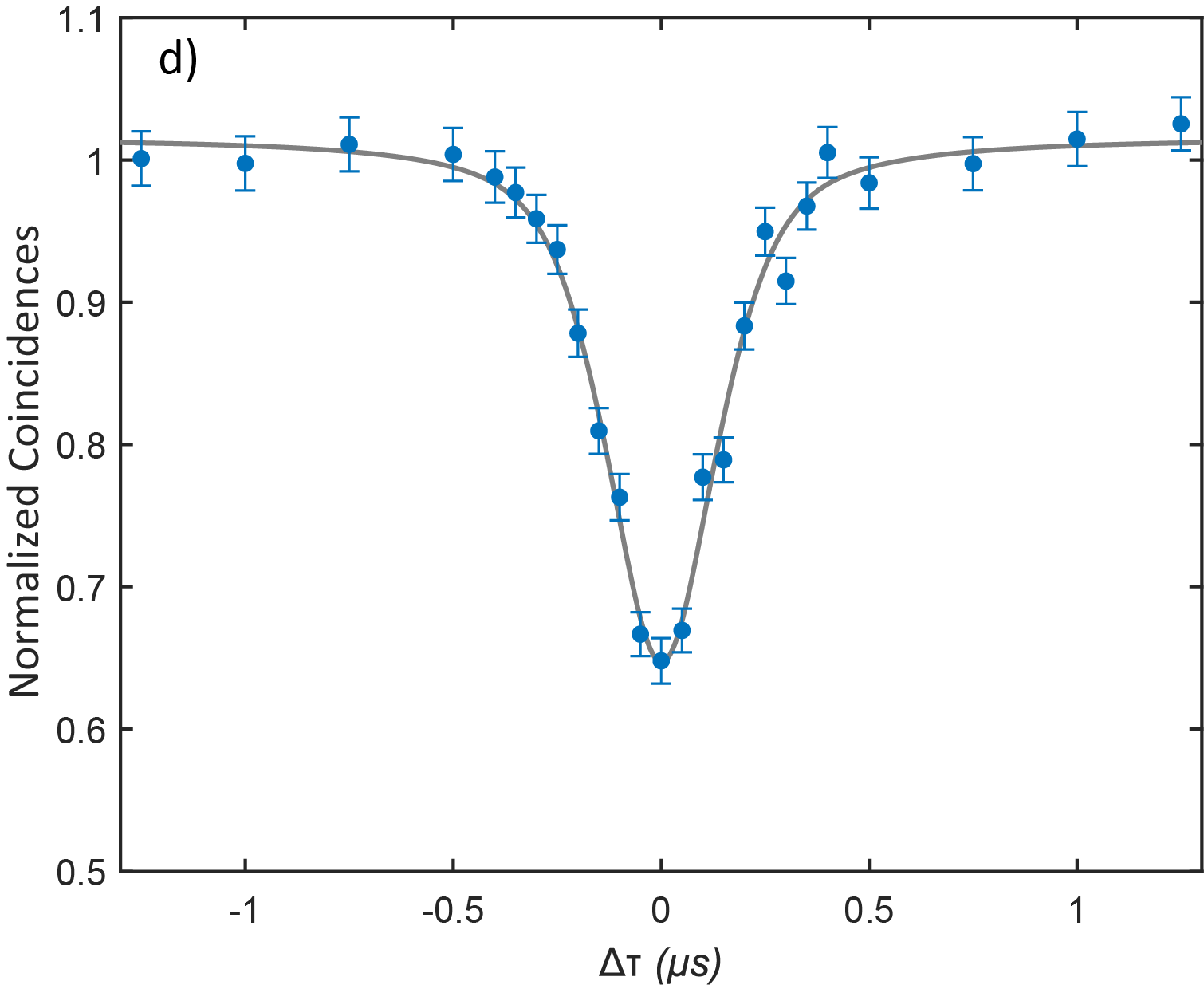}
 \includegraphics[width=0.32\linewidth]{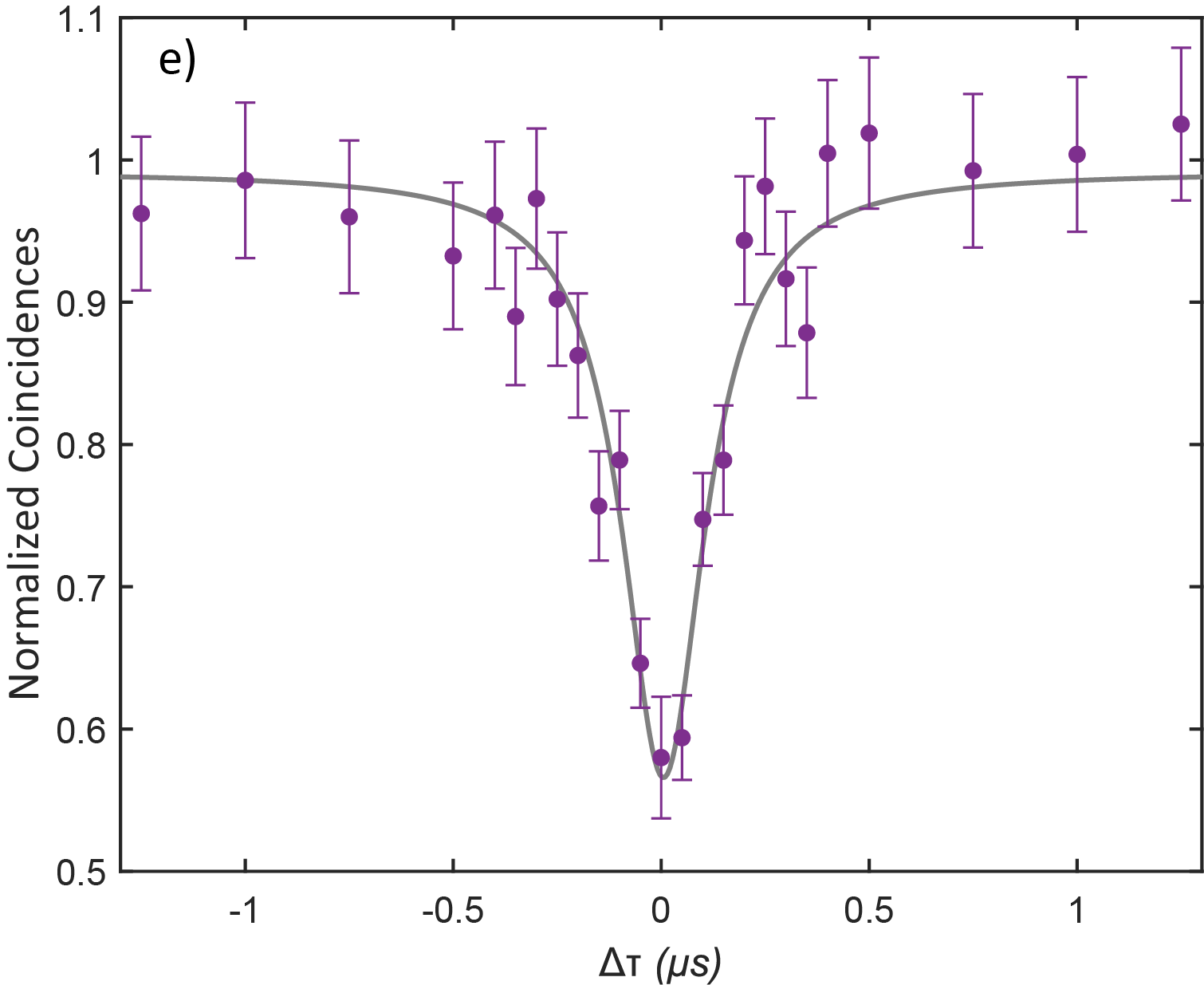}
 \includegraphics[width=0.32\linewidth]{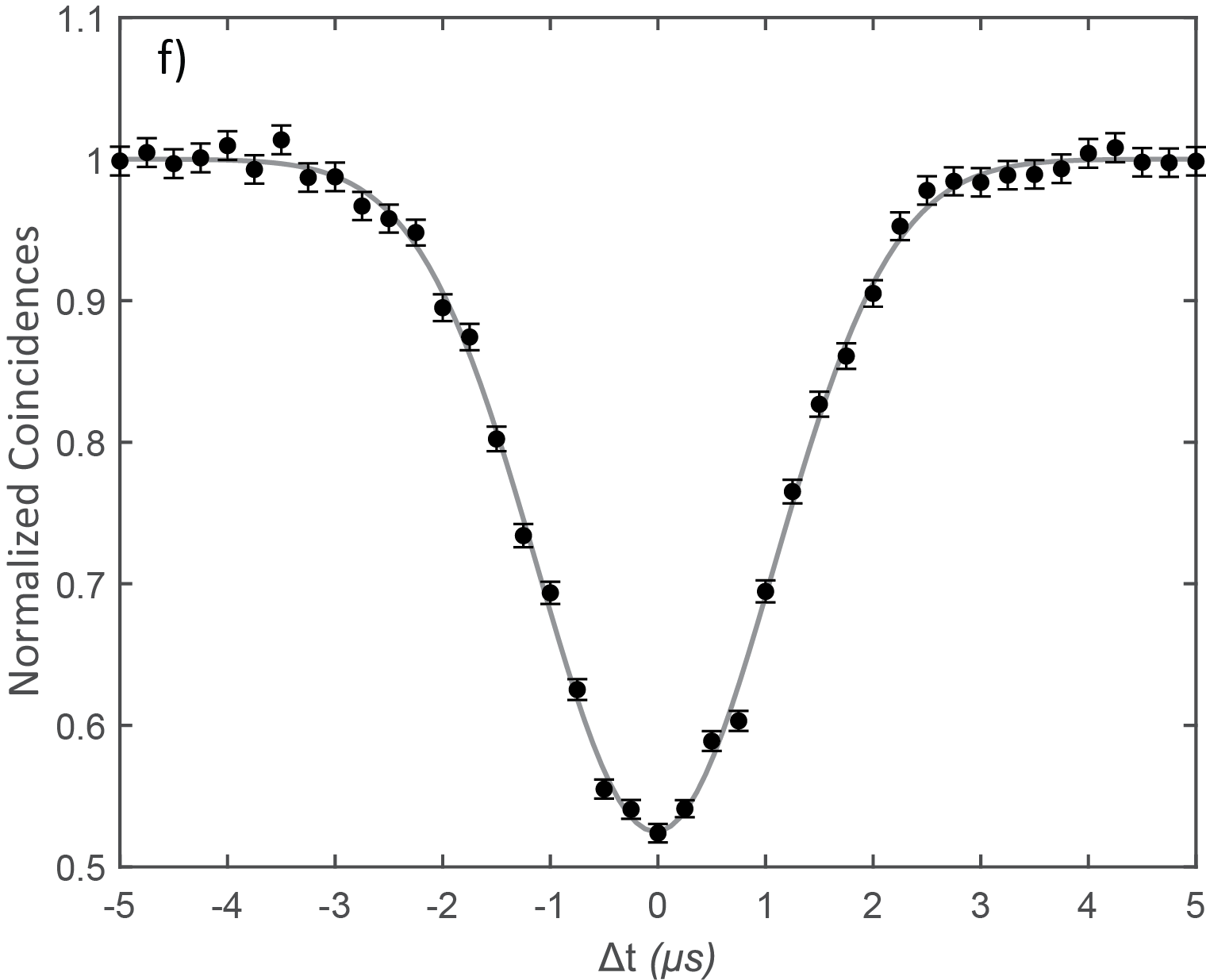}

 \caption{\textbf{Individual HOM interference plots with error bars for different estimated signal-to-background ratios, shown in Fig. \ref{fig:visibilitymodel}}:  a) The combination of subplots a-e, without error bars, as Fig. 5a. b) Coincidences obtained using a region of interest (ROI) of $4.023 \mu s$, with a signal-to-background ratio (SBR) of $1.6$ and a calculated visibility of $17.1 \pm 0.8 \%$. c) Coincidences obtained using a ROI of $1.279 \mu s$, and a SBR of $5.0$ and a calculated visibility of $31.6 \pm 1.1 \%$. d) Coincidences obtained using a ROI of $0.638 \mu s$, a SBR of $7.8$ and a calculated visibility of $36.2 \pm 1.3 \%$. e) Coincidences obtained using a ROI of $0.157 \mu s$, a SBR of $11.9$, and a calculated visibility of $42.9 \pm 3.4 \%$. f) HOM interference between pulses when no memories were involved and resulted in a visibility of $48.5 \pm 0.5 \%$. }
 \label{fig:HOMDipsGenII}
\end{figure*}

For the HOM experiment in section II, with stored qubits and varying polarization, we post-select a Region of Interest (ROI), which defines the coincidence window beginning at the qubit retrieval time and equal in duration to the FWHM of the input pulses, i.e. $\rm 0.4 \mu s $. The interference visibility is defined as $\rm V_{HOM} = 1 - C_{coin}/(S_1 S_2)$ where $\rm C_{coin}$ is the probability of two-detector coincidence and $S_{1,2}$ are the probabilities of detecting an event at each detector. For the polarization degree of freedom, $\rm V_{HOM} \propto (\cos{\theta})^2$, where $\theta$ is the relative polarization angle between two weak coherent pulses \cite{Moschandreou2018}. We plot the polarization angles vs. coincidence rate $\rm C_{coin}/(S_1 S_2)$ with $\rm C_{coin}/(S_1 S_2) = N C_{12}/(C_1C_2)$, where $C_{12}$ is the number of coincidences, $\rm C_{1(2)}$ are the counts at each detector, and $\rm N$ is the total number of triggers. When the two pulses are distinguishable, we expect the coincidence rate to be unity, i.e. $\rm C_{coin}/(S_1 S_2) = 1$. The visibility is obtained by fitting $\rm C_{12}/(C_1 C_2) = A \left(1-V_{HOM}\cos^2(\theta-\theta_0)\right)$ to the plot via a maximum likelihood method (MLM). \\

For the HOM experiment section III using stored pulses in the tabletop memories with varying delay, the separate ROIs of width $\rm 0.3 \mu s$ are set beginning at the time of retrieval of each pulse. As one of the two pulses is delayed (Alice), the corresponding ROI is also shifted accordingly. The total width of the ROI is fixed at $\rm 0.6 \mu s$. When the ROIs of two pulses overlap, an additional ROI is added in the dark count-only region to maintain a constant ROI width and thus a constant dark-count rate. The interference visibility was obtained by fitting the normalized coincidence rate to a Gaussian function via MLM.\\

Compared to sections II and III where the ROI is kept constant, ROIs of various lengths are defined in order to create plots in Fig. \ref{fig:visibilitymodel} for the HOM experiment of pulses stored in the Gen II portable memories. The signal-to-background is varied while processing the data by varying the ROI width. Changing the ROI changes the signal-to-background ratio since the background is uniformly distributed while the signal is concentrated in a short pulse. Picking a very narrow(wide) window around the signal pulse results in a high(low) SBR. This lets us see the effect of the signal-to-background ratio on system performance.\\

In addition, during the data-taking process, fluctuations in the transmission through the system can impact the signal-to-background ratio if the ROI timing considered is kept constant. In order to circumvent the problem, the lengths of ROIs were defined as a fraction of the peak retrieval height while completely overlapped ($\Delta \tau = 0$). This was in an effort to keep the signal-to-background ratio more consistent under any fluctuations during the experiment as compared to a fixed ROI time. These ROIs were shifted according to the difference in storage times to follow the retrieval peaks. As before, an additional ROI is added in the dark count-only region to maintain a constant ROI width for each analysis set while the ROIs of two pulses overlap. The interference visibility was obtained by fitting the normalized coincidence rate to a Voigt function via MLM (Fig. \ref{fig:HOMDipsGenII}b-e). \\

Table \ref{tab:ROI} lists the different ROI windows and the corresponding signal-to-background ratios obtained. Fig. \ref{fig:HOMDipsGenII} shows $g^{(2)}$ plots obtained from the same data set using different ROI widths. As the ROI is narrowed from plot (a) to (e) in Fig. \ref{fig:HOMDipsGenII}, we see an increase in signal-to-background ratio accompanied by an increase in visibility.\\

\bibliography{main}
\bibliographystyle{apsrev4-1}

\end{document}